\documentclass[a4paper,12pt]{article}

\pdfoutput=1

\usepackage{graphicx}
\usepackage{psfrag}
\usepackage{epsfig}
\usepackage{colordvi}
\usepackage{amsbsy}
\usepackage{amssymb}
\usepackage{amsmath}
\usepackage{amsfonts}
\usepackage{exscale}
\usepackage{fancyhdr}
\usepackage{color}
\usepackage{pslatex} 
\usepackage{palatino}
\usepackage{newcent}
\usepackage{scalefnt}
\usepackage{authblk}
\usepackage[utf8]{inputenc}
\usepackage{setspace}
\usepackage[english]{babel}

\usepackage{natbib}
\setcitestyle{authoryear,open={(},close={)}}

\AtBeginDocument{\RequirePackage{lmodern, times}}

\date{}

\newcommand{\range}[2]{{#1:#2}}
\newcommand{\md}{\mathrm{d}}

\begin{document}

\pagestyle{fancy}

\thispagestyle{empty}

\title{The reconstruction of sound speed in the Marmousi model by the boundary control method}

\author{I.~B.~Ivanov$^{1,2}$, M.~I.~Belishev$^{1,3}$, V.~S.~Semenov$^1$\\
\vspace{0.25cm}
{\small
$^1$ St.~Petersburg State University, Institute of Physics, St. Petersburg, Russia\\
$^2$ St.~Petersburg Nuclear Physics Institute, Theoretical Physics Division, Gatchina, Russia\\
$^3$ St.~Petersburg Department of V.A. Steklov Institute of Mathematics, St. Petersburg, Russia\\
emails: contact@ivisoft.org, m.belishev@spbu.ru, sem@geo.phys.spbu.ru
}
}

\maketitle

\label{firstpage}

\begin{abstract}
We present the results on numerical testing of the Boundary
Control Method in the sound speed determination for the
acoustic equation on semiplane. This method for solving
multidimensional inverse problems requires {\it no a priory information}
about the parameters under reconstruction.
The application to the realistic Marmousi model demonstrates that the
boundary control method is workable in the case of complicated and irregular field of
acoustic rays. By the use of the chosen boundary controls, an
`averaged' profile of the sound speed is re\-co\-ve\-red (the
relative error is about $10-15\%$). Such a profile can be further
utilized as a starting approximation for high resolution iterative
reconstruction methods.
\end{abstract}

{\bf Key words:} acoustics, inverse problem, modelling

\section{Introduction}

\label{sec:intro}

The Boundary Control Method (BCM) is {\it ab initio} approach to
multidimensional inverse problems based on ideas and results of
control theory, asymptotic methods (for propagation of
discontinuities), functional analysis, and geometry: see
\citep{belishev-1987, belishev-1997, belishev-2007} and references
therein. Its simple and clear mathematical background makes the
method of rather general scope. In particular, the BCM is
developed for acoustics \citep{belishev-1987}, electrodynamics
\citep{belglas-2001}, Shr\"odinger equation \citep{avdbel-2004}, 1D
Dirac system \citep{belmik-2014}, 1D two-velocity system
\citep{beliva-2005}, reconstruction of Riemannian manifolds
\citep{belishev-1997, beldem-2011}, metric graphs
\citep{belvak-2006}, and other dynamic systems.

The BCM provides {\it time optimal} step-by-step reconstruction
procedure, which requires {\it no ad hoc assumptions} about the sound
speed profile. It works in the case of data given {\it on a part}
of the domain boundary. Variants of BCM are developed and
numerically tested for different problems in \citep{belgotiva-1997,
belgot-1999, pesbolkaz-2010, pesboldan-2012, pes-2012, pes-2014,
oks-2013}.

There exist another (not optimal) direct reconstruction methods
which are numerically (and experimentally) tested: see
\citep{kabshish-2004, kabsatshish-2005, kabshish-2011,
belklib-2008, belklib-2012} and references therein. A time optimal
and {\it data optimal} approach by V.~G.~Romanov \citep{rom-1996} is
not implemented and tested yet.

The goal of this paper is to provide a concise and transparent
outline of dynamic variant of BCM for the acoustic equation and
discuss recent numerical results on the sound speed reconstruction
in several test cases. A rigorous and detailed exposition of the
method can be found in the reviews \citep{belishev-1997,
belishev-2007}. Previous results on numerical testing of the same
BCM version for the wave equation are published in
\citep{belgotiva-1997, belgot-1999} and \citep{belivsem-2015}.

We consider a dynamic system governed by the acoustic equation in
a domain $\Omega \subset{\bf R}^n$ ($n \geq 2$) with the boundary
$\Gamma$:
\begin{align}
\label{acoustic 1} &\left[\partial_t^2 - c^2\Delta\right] u = 0
&& \rm in\,\,\Omega \times (0,T)\\
\label{acoustic 2} &u\big|_{t=0} = \partial_t
u\big|_{t=0} = 0 && \rm in\,\,\Omega\\
\label{acoustic 3} & u\big|_{\Sigma^T} = f.
\end{align}
Here $t$ is a time, $x = (x^1, \dots, x^n)\in {\bf R}^n$ are Cartesian coordinates, $c=c(x)
> 0$ is a speed of sound, $\Sigma^T:=\Gamma\times[0,T]$.
A solution ({\it wave}) $u = u^f(x,t)$ describes a perturbation of
the acoustic pressure in the medium caused by a boundary source
({\it control}) $f$ acting from $\Gamma$ during the probing time
$T$. An `input $\rightarrow$ output' correspondence of the system
is realized by a {\it response operator} $R^T:f
\mapsto\partial_\nu u^f|_{\Sigma^T}$ ($\nu$ is the outward normal
to $\Gamma$).

Let $R^{2T}$ be the response operator of the same  system
(\ref{acoustic 1})--(\ref{acoustic 3}) with  the double probing
time $2T$. Due to the finiteness of the wave propagation speed,
operator $R^{2T}$ is determined by the values of $c$ in the
near-boundary subdomain $\Omega^T$ filled up with waves at the
moment $t = T$,  and does not depend on the behavior of $c$ in
$\Omega\backslash\Omega^T$. Such a local character of the
dependence motivates the following  relevant statement of the
inverse problem: {\it given $R^{2T}$ recover the speed $c$ in the
subdomain $\Omega^T$}.

To solve this problem, the BCM exploits certain subtle properties
of the waves. The principal role is played by a completeness of
waves, which is interpreted in control theory as a boundary
controllability of the system (\ref{acoustic 1})--(\ref{acoustic
3}). Also, the geometrical optics is used for extracting
information about the medium from the wave field jumps propagating
through $\Omega$ and being detected on $\Gamma$. The crucial point
is that, for the given controls $f, g$, the integrals of the form
$\int_\Omega\md x\,c^{-2} u^f u^g$ are determined by the inverse data, i.e.,
can be computed via operator $R^{2T}$. These facts enable the
external observer to recover the {\it images} of waves on the
space-time surface $\Sigma^T$. Roughly speaking, an image is just
a wave written in the {\it ray coordinates}, which are associated
with the acoustic rays orthogonal to $\Gamma$. Along with the wave
images, the observer can recover images of the Cartesian
coordinate functions and, thus, determine the connection between
the Cartesian and ray coordinates. Such a connection easily
determines the sound speed $c$ in $\Omega^T$ that solves the
inverse problem. In the paper, we propose the procedure, which
recovers $c$ in a ray tube $B^T_\sigma \subset \Omega$ covered by
acoustic rays, which emanate from a part $\sigma \subset \Gamma$.
In capacity of the inverse data, the response $R^{2T}f$ on the
controls $f$ acting on $\sigma$, is used.

The paper is organized as follows. In the first three sections we
introduce the reader to main notions and known results, which form
a 'language` of the boundary control method. In
Section~\ref{sec:geom} we describe a convenient and natural system
of semigeodesic coordinates. In Section~\ref{sec:bc} the notions
of control theory are applied to the dynamic system (\ref{acoustic
1})--(\ref{acoustic 3}). In Section~\ref{sec:optics} we use the of
geometric optics formulas for propagation of discontinuities to
derive the amplitude formula, which is a main computational device
of BCM. In Section~\ref{sec:recon} we combine all findings and
provide step-by-step reconstruction procedure for the speed of
sound. Finally, in Section~\ref{sec:test} some numerical results
of application of BCM to several test cases are presented and
discussed.

\section{Geometry}

\label{sec:geom}

\subsection{Metric}

The speed of sound $c$  determines a {\it $c$-metric} in $\Omega$
with the distance between points
\begin{align}
\label{cmetric} \tau(x,y) := \text{inf} \int\limits_x^y\,\frac{\md
l}{c}\,,
\end{align}
where the infimum is taken over all smooth curves connecting $x$
and $y$, $\md l$ is a length element in ${\bf R}^n$. In dynamics,
$\tau(x,y)$ coincides with the travel time needed for a wave
initiated at $x$ to reach $y$. The geodesics of the $c$-metric are
the curves realizing the infimum in (\ref{cmetric}).

Let $\sigma\subseteq\Gamma$ be an  open set at  the boundary. In
what follows, unless otherwise specified, we assume that $\sigma$
is fixed.  The travel time needed for a wave initiated on $\sigma$
to reach a point $x\in\Omega$ is called an {\it eikonal}:
\begin{align}
\label{eikonal} \tau_\sigma(x) := \min_{y\in\sigma}\,\tau(y,x).
\end{align}
Fix a $\xi \geq 0$. The level set of the eikonal
\begin{align}
\Gamma^\xi_\sigma:=\left\{x\in\Omega \mid \tau_\sigma(x) =
\xi\right\}
\end{align}
is a wave front surface that bounds (together with $\Gamma$) a
subdomain filled by waves, which move into $\Omega$ from $\sigma$,
at $t=\xi$,
\begin{align}
\Omega^\xi_\sigma:=\left\{x\in\Omega \mid \tau_\sigma(x) <
\xi\right\}.
\end{align}

In the case $\sigma=\Gamma$ we simplify the notation and write
$\tau,\, \Gamma^\xi, \,\Omega^\xi$.

\subsection{Ray coordinates}
\begin{figure}[h]
\begin{center}
\includegraphics[width=0.7\columnwidth]{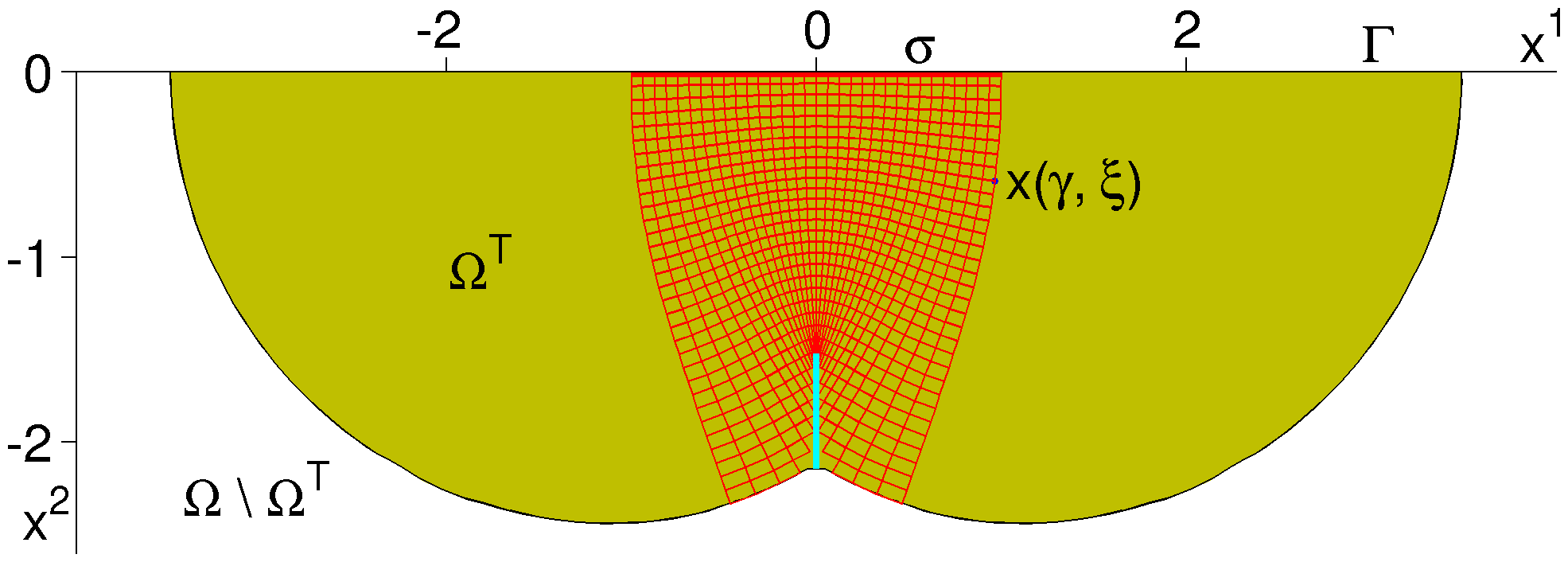}
\end{center}
\caption[]{A typical simple geometry of rays on semiplane:
domain $\Omega^T$ (fill), semigeodesic coordinates $x(\gamma,\xi)$ (mesh), cut locus (vertical line).
The speed of sound $c(x)$ is taken from Section \ref{sec:test:case_1} with $\sigma=\{x\in\Gamma \mid |x^1| < 1\}$, $T = 2.5$.}
\label{fig:sgc}
\end{figure}

Fix a point $\gamma\in\sigma$. Let $x(\gamma,\xi)$ be the endpoint
of the geodesic curve in $c$-metric ({\it ray}) starting from
$\gamma$ orthogonally to $\sigma$ and parametrized by its
$c$-length $\xi$. For $T>0$, all such rays starting from $\sigma$
cover a subdomain ({\it tube})
\begin{align}
\nonumber B^T_\sigma:=\bigcup_{\gamma\in\sigma}\bigcup_{0\le\xi\le
T} x(\gamma,\xi).
\end{align}
We say $\sigma$ to be a bottom of the tube. A typical picture of
the ray field in  the case of  the semiplane is shown in
Fig.~\ref{fig:sgc}, here $\Omega=\{x \mid x^2 < 0\}$, $\Gamma=\{x
\mid x^2 = 0\}$, $\sigma=\{x\in\Gamma \mid |x^1| < 1\}$, and speed
of sound $c(x)$ is taken from Section \ref{sec:test:case_1}.

If $T$ is sufficiently small then the ray family is regular: only
one ray passes through any point in $B^T_\sigma$. In this case, to
each $x \in B^T_\sigma$ we preassign the pair $(\gamma, \xi)$ such
that $x=x(\gamma,\xi)$. This pair is called the semigeodesic ({\it
ray}) coordinates of $x$.

The connection between the ray and Cartesian coordinates can be easily calculated
by solving the Euler-Lagrange equations for functional (\ref{cmetric}): if
$x=x(\gamma,\xi)=\left(x^1(\gamma,\xi), \dots ,
x^n(\gamma,\xi)\right)$ then
\begin{align}
\nonumber \frac{\partial x^k}{\partial\xi} = c^2\,v^k, \quad
\frac{\partial v^k}{\partial\xi} = - \frac{1}{c}\,\frac{\partial
c}{\partial x^k}\,, \qquad \xi>0
\end{align}
with the initial data $x(\gamma,0)=\gamma$ and $v(\gamma,0) =
-\nu/c$.

For large enough $T$'s, the ray field in the tube may loose
regularity. In particular, the so called multiple points, which
are connected with $\sigma$ through more than one ray, may appear.
A set
$$
\omega_\sigma :=\overline{\{x \in B^T_\sigma\,|\,\, x\,\, {\rm is\,\,\, multiple}\}}
$$
(the bar means a closure in ${\bf R}^n$) is said to be a
separation set ({\it cut locus}). The remarkable fact is that the
cut locus is `small': its volume equals zero. Therefore the
semigeodesic coordinates can be used almost everywhere in
$B^T_\sigma$. In particular, if $\sigma=\Gamma$, one can use them
almost everywhere in $\Omega$.

\subsection{Screen and images}
The important characteristic of the ray field is its divergence,
which plays a role of amplitude factor in geometric optics
formulas for propagation of wave discontinuities. Fix a point
$\gamma\in\sigma$ and denote $\sigma_\varepsilon(\gamma, 0)$ the
intersection of $\sigma$ with a ball of small radius $\varepsilon$
centered at $\gamma$. Consider a tube $B^T_\varepsilon(\gamma)$
with the bottom $\sigma_\varepsilon(\gamma, 0)$, and let
$\sigma_\varepsilon(\gamma, \xi) :=
B^T_{\varepsilon}(\gamma)\cap\Gamma^\xi_\sigma$.
The function of the ray coordinates
\begin{align}
\label{raydiv}
J(\gamma,\xi) := \lim_{\varepsilon\rightarrow
0}\,\frac{\text{mes}\, \sigma_\varepsilon(\gamma,\xi)}{\text{mes}\,\sigma_\varepsilon(\gamma, 0)}
\end{align}
is called a {\it divergence} at point $x(\gamma,\xi)$. Here $\rm
mes$ is the Euclidean surface measure (area) in ${\bf R}^n$.

The divergence determines a function
$$
\beta(\gamma,\xi):=
\left[\frac{J(\gamma,\xi)J(\gamma,0)}{c(x(\gamma,\xi))c(x(\gamma,0))}\right]^{\frac{1}{2}}\,,
$$
which is related with the Jacobian of the passage from the ray to
Cartesian coordinates \citep{babbul-1991}.

Let $f(x)$ be a function given in the tube $B^T_\sigma$. A
function $\tilde f$ of the ray coordinates defined on a {\it
screen} $\Sigma^T_\sigma := \sigma\times[0,T]$ by
\begin{align}
\label{image} \tilde
f(\gamma,\xi):=\beta(\gamma,\xi)\,f(x(\gamma,\xi))\,, \qquad (\gamma,\xi) \in \Sigma^T_\sigma
\end{align}
is called an {\it image} of $f$. This definition will be motivated later in section \ref{sec:ampfor}.

\section{Boundary controllability}

\label{sec:bc}

\subsection{Spaces and operators}

The dynamic system (\ref{acoustic 1})--(\ref{acoustic 3}) can be
attributed with spaces and operators as it is customary in control
theory.

The space $\mathcal{F}^T := L_2(\Sigma^T)$ of the square-summable
boundary controls with the inner pro\-duct
\begin{align}
\label{outer} \left(f,g\right)_{\mathcal{F}^T} :=
\int_{\Sigma^T}{\md \Gamma}\,{\md t}\,f(\gamma,t)\,g(\gamma,t)
\end{align}
($\md\Gamma$ is the Euclidean surface element on $\Gamma$) is
called an {\it outer space} of the system. It contains an
increasing family of subspaces
\begin{align}
\nonumber \mathcal{F}^{T,\xi} := \left\{f\in \mathcal{F}^T \mid
f(\cdot,t)=0, \,\,\,0\le t\le T-\xi\right\},\quad  0\le \xi\le T
\end{align}
formed by the delayed controls. Also, we deal with the subspaces
of controls acting from $\sigma$:
 \begin{align}
\nonumber \mathcal{F}^{T,\xi}_\sigma := \left\{f\in
\mathcal{F}^{T,\xi} \mid \text{supp}\,f \subset
\Sigma^T_\sigma\right\},\qquad 0\le \xi\le T\,.
\end{align}

The space of waves ${\mathcal{H}} := L_2(\Omega; c^{-2}\md x)$
with the  product
\begin{align}
\label{inner} \left(v,w\right)_{{\mathcal{H}}} :=
\int\limits_\Omega \md x\,c^{-2}(x)\,v(x)\,w(x)
\end{align}
is said to be an {\it inner space}. It contains a family of
subspaces
\begin{align}
\nonumber \mathcal{H}^{\xi}_\sigma := \left\{v\in{\mathcal{H}}^T
\mid \text{supp}\,v\subseteq
\overline{\Omega}^\xi_\sigma\right\},\qquad 0\le \xi\le T\,.
\end{align}
Since the waves propagate with the speed $c$, for any control $f
\in \mathcal{F}^{T,\xi}_\sigma$ the corresponding wave
$u^f(\cdot,T)$ is supported in the subdomain $\Omega^\xi_\sigma$,
so that $u^f(\cdot,T)$ turns out to be an element of the subspace
$\mathcal{H}^{\xi}_\sigma$.

An `input $\rightarrow$ state' correspondence in the system is
described by the {\it control operator}
$W^T:\mathcal{F}^T\rightarrow {\mathcal{H}}$, $$W^T f :=
u^f(\cdot,T)\,.$$ By the above mentioned finiteness of the wave
propagation speed, the relation
\begin{align}
\nonumber
W^T\mathcal{F}^{T,\xi}_\sigma\subseteq\mathcal{H^\xi_\sigma}\,,
\qquad 0\le \xi\le T
\end{align}
is valid.

An `input $\rightarrow$ output' correspondence is realized by the
{\it response operator} $R^T:\mathcal{F}^T\rightarrow
\mathcal{F}^T$,
 $$
R^T f := \partial_\nu u^f\big|_{\Sigma^T}
 $$
defined on the smooth enough controls vanishing at $t=0$. Let us
notify: if a control $f$ acts from $\sigma$, the response $R^Tf$
is assumed to be observed on the {\it whole} boundary $\Gamma$
(not only on $\sigma$). However, by the finiteness of $c(x)$, such
a response is supported on the part $\Gamma \cap \Omega^T_\sigma$,
i.e., vanishes far from $\sigma$.

A {\it connecting operator} $C^T:\mathcal{F}^T\rightarrow
\mathcal{F}^T$,
 $$
C^T:=(W^T)^\ast W^T
 $$
relates the metrics of the inner and outer spaces: for any $f,
g\in \mathcal{F}^T$, one has
\begin{align}
\label{Cop} (u^f(\cdot,T), u^g(\cdot,T))_{\mathcal{H}} = (W^T f,
W^Tg)_{\mathcal{H}} = (C^T f, g)_{\mathcal{F}^T}.
\end{align}

\subsection{Wave products}

The role of the connecting operator
stands out due to the following remarkable fact: the result of
action of $C^T$ can be expressed in explicit form via the
response operator as follows. 

Take a control $f \in \mathcal{F}^T$. Extend it from the time
interval $(0,T)$ to $(0,2T)$ by oddness with respect to $t=T$:
\begin{align}
\nonumber
f_-(\gamma,t) :=
\left\{
      \begin{array}{rc}
        f(\gamma,t), & \quad 0 < t < T\\
       -f(\gamma,2T-t), &\quad T< t< 2 T
    \end{array}
\right\}.
\end{align}
Then define a `double control'
$$
F(\gamma,t)\,:=\,\frac{1}{2}\,\int\limits_0^t \md t' \,f_-(\gamma, t')\,, \qquad 0<t<2T\,.
$$

Let $u^F$ be a solution to the system (\ref{acoustic 1})--(\ref{acoustic 3}) {\it with the final moment} $t=2T$.
As it is shown in \citep{belishev-1997}, the equality
\begin{align}
\nonumber & (C^T f)(\gamma,t) = \partial_\nu
u^F(\gamma,t) - \partial_\nu u^F(\gamma,2T-t)\,=\\
\label{C^T via R^2T}&=\,\left(R^{2T}F\right)(\gamma,t) -
\left(R^{2T}F\right)(\gamma,2T-t)\,, \qquad 0\leq t\leq T
\end{align}
holds, where $R^{2T}$ is the corresponding response operator.

The external observer operates at the boundary, can set controls
and create waves into the domain but he/she cannot see the waves
themselves. Nevertheless, by (\ref{Cop}) and (\ref{C^T via R^2T})
the observer is able to determine inner products of these
invisible waves through the measurements on the boundary!

As a consequence, for any family of controls $\{f_\alpha\} \subset
\mathcal{F}^T$, one can determine the {\it Gram matrix} of the
corresponding waves $u^{f_\alpha}(\cdot,T)$:
\begin{align}
\notag & A_{\alpha\beta}:=
\left(u^{f_\alpha}(\cdot,T),u^{f_\beta}(\cdot,T)\right)_{\mathcal{H}}\overset{(\ref{Cop})}=
\left(C^T{f_\alpha},{f_\beta}\right)_{\mathcal{F}^T}\overset{(\ref{C^T via R^2T})}=\\
\label{Gram general}& = \int\limits_{\Sigma^T} \md t\, \md
\Gamma\, \left[\left(R^{2T}F_\alpha\right)(\gamma,t) -
\left(R^{2T}F_\alpha\right)(\gamma,2T-t)\right]\,f_\beta(\gamma,t)\,.
\end{align}

One more option is to determine the products of waves and harmonic
functions. Let a function $a=a(x)$ satisfy $\Delta a=0$ in
$\Omega^T$. A simple integration by parts provides the equality
\begin{align}
\label{(a, u^f)}
\left(a,u^{f}(\cdot,T)\right)_{\mathcal{H}} = \int\limits_{\Sigma^T} \md t\, \md
\Gamma\,(T-t)\,\left[a(\gamma)(R^Tf)(\gamma,t)-\partial_\nu a(\gamma)\,f(\gamma,t)\right]\,,
\end{align}
which represents the product via the response operator, see \citep{belishev-1997}.

\subsection{Dual system}

The dynamic system
 \begin{align}
\label{dual 1} &\left[\partial_t^2 - c^2\Delta\right] v = 0 && \rm in\,\,\Omega \times (0,T)\\
\label{dual 2} &v\big|_{t=T} = 0, \quad \partial_t
v\big|_{t=T} = y &&  \rm in\,\,\Omega\\
\label{dual 3} & v\big|_{\Sigma^T} = 0
\end{align}
with $y \in {\mathcal{H}}$ is called {\it dual} to the system
(\ref{acoustic 1})--(\ref{acoustic 3}). Its solution $v=v^y(x,t)$
describes a wave, {which is produced by a velocity perturbation
$y$ and propagates (in the reversed time) in the domain with the
rigidly fixed boundary. The value $\partial_\nu v^y|_{\Sigma^T}$
is proportional to a force, which appears as a result of
interaction between the wave and the boundary at a point $\gamma$
at a moment $t$.

It is useful to introduce an {\it observation operator} $O^T :
{\mathcal{H}}\to \mathcal{F}^T$ associated with the dual system
and mapping the perturbation $y$ to the force observed at the
screen $\Sigma^T$,
\begin{align}
\label{obsop} O^T y := \partial_\nu v^y\big|_{\Sigma^T}.
\end{align}

There is an important relation between solutions $u^f$ and $v^y$
that motivates the term `dual'. For any boundary control
$f\in\mathcal{F}^T$ and perturbation $y\in\mathcal{H}$ the
following equality holds:
\begin{align}
\label{duality} (u^f(\cdot,T), y)_{\mathcal{H}} = (f,
\partial_\nu v^y)_{\mathcal{F}^T}.
\end{align}
By the definition of the operators, it is equivalent to $(W^T f,
y)_{\mathcal{H}} = (f, O^T y)_{\mathcal{F}^T}$ that implies
$(W^T)^\ast = O^T$. Hence, for the connecting operator (\ref{Cop})
we get
\begin{equation}
\label{C^T=O^T W^T}
 C^T\, = \,O^T W^T\,.
\end{equation}

\subsection{Wave bases}

\label{sec:wavbas}

Let us choose a function $y$ supported in subdomain
$\Omega^T_\sigma$. Consider a {\it boundary control problem}: find
a control $f$ acting from $\sigma$ such that the wave $u^f$
satisfies
\begin{align}
\nonumber u^f(\cdot,T)\, = \,y\,.
\end{align}
In other words, the question is whether we can manage the shape of
waves from the boundary. The answer is the following. For any
$y\in\mathcal{H}^T_\sigma$ and arbitrarily small $\varepsilon > 0$
one can find a control $f\in\mathcal{F}^T_\sigma$ such that
\begin{align}
\nonumber \|y-u^f(\cdot,T)\|_{\mathcal{H}}^2 =
\int\limits_{\Omega^T_\sigma} \md x\,c^{-2}(x)\,|y(x)-u^f(x,T)|^2
< \varepsilon.
\end{align}
In control theory, this property of system (\ref{acoustic
1})--(\ref{acoustic 3}) is referred to as an {\it approximate
boundary controllability}. It means that the set of waves produced
from the boundary is reach enough to approximate functions. In
other terms, the set $\{u^f(\cdot,T)\,|\,\,f \in
{\mathcal{F}^T_\sigma}\}$ is {\it complete} in
${\mathcal{H}^T_\sigma}$.

Controllability is the fact of affirmative character for inverse
problems: the very general principle of system  theory claims that
the richer is the set of states of a dynamical system, which the
external observer can create by means of controls, the richer is
information about the system which the observer can extract from
the external measurements.

One more important consequence of controllability is the existence
of wave bases. Let $\{f_\alpha\}\subset \mathcal{F}^T_\sigma$ be a
complete and linearly independent family of controls. A
completeness means that any $f \in \mathcal{F}^T_\sigma$ can be
approximated by the family elements, i.e., one can expand
$
f\approx \sum \limits_{\alpha=1}^N\lambda^N_\alpha f_\alpha
$
with arbitrary precision by the proper choice of the coefficients
$\lambda^N_\alpha$ and a finite $N$. Owing to the controllability,
the corresponding family of waves
$u_\alpha:=u^{f_\alpha}(\cdot,T)$ turns out to be complete in
$\mathcal{H}^T_\sigma$, i.e., any $y\in \mathcal{H}^T_\sigma$ can
be approximated as
 \begin{align}
\label{wavexp} y \,\approx\, \sum \limits_{\alpha=1}^N
c^N_\alpha\, u_\alpha\,.
\end{align}
It is the family $\{u_\alpha\}$ which we call a {\it wave basis}.

As it is well known, the optimal {\it least-squares} approximation in (\ref{wavexp}) is
provided by the coefficients $c^N_\alpha$ which satisfy the Gram linear system
 \begin{align}
\label{Gram system} \sum \limits_{\alpha=1}^N A_{\beta\alpha}c^N_\alpha\,=\,b_\beta,\quad
A_{\alpha\beta} = (u_\alpha, u_\beta)_{\mathcal{H}},\quad
b_{\beta} = (y, u_\beta)_{\mathcal{H}},\quad
\beta=1,\dots N.
\end{align}
For the given family $\{f_\alpha\}$, the matrix entries can be
found via the response operator: see (\ref{Gram general}). Also,
if $y=a(x)$ is harmonic, one can determine $b_\beta$ by (\ref{(a,
u^f)}). Such an option will play a crucial role in solving the
inverse problem.
%

\section{Geometric optics}

\label{sec:optics}

\subsection{Propagation of discontinuities}

The well known fact is that discontinuous controls generate
discontinuous waves. The discontinuities of waves propagate along
the acoustic rays, and their jumps can be expressed in explicit
form by means of formulas of geometric optics.

Choose on $\sigma$ a control $f$ and fix a parameter $\xi\in (0,
T)$. Consider the discontinuous control
\begin{align}
\nonumber
f_\xi(\gamma,t) :=
\left\{
      \begin{array}{rr}
       0, & \quad 0\leq t < T-\xi\\
       f(\gamma,t), & \quad T-\xi\leq t\leq T
    \end{array}
    \right\},
\end{align}
which has a jump $f(\gamma,T-\xi)$ at $t=T-\xi$. This jump
initiates a jump of the wave $u^f$, which propagates along the
rays into the domain. At the final moment $t=T$ the wave jump is
placed on the forward front surface $\Gamma^\xi_\sigma \cap
B^T_\sigma$, its amplitude being equal to
\begin{align}
\label{discont}  \lim_{\tau\rightarrow\xi-0}
u^{f_\xi}(x(\gamma,\tau),T) =
\left[\frac{c(x(\gamma,\xi))J(\gamma,0)}{c(x(\gamma,0))J(\gamma,\xi)}\right]^{\frac{1}{2}}f(\gamma,T-\xi).
\end{align}

Such relations are well known as the geometric optics formulae,
see e.g. \citep{babbul-1991}. As they show, up to a geometric
factor $[...]^{\frac{1}{2}},$  the wave jump reproduces the shape
of the control jump. Let us remark that (\ref{discont}) is valid
only in the `regular' part of the tube $B^T_\sigma$, i.e., outside
the cut locus.

\subsection{Jumps in dual system}
Assume $\sigma$ and $T$ to be such that the ray field is regular
in the tube $B^T_\sigma$. Take a smooth function
$y\in\mathcal{H}^T_\sigma$ and fix a $\xi \in (0,T)$.
Let $y^\xi$ and $y^\xi_\perp$ be the truncated functions defined by
\begin{align}
\label{projector}
y^\xi(x) := 
\left\{
      \begin{array}{rl}
       y(x), & \quad x\in\Omega^\xi\\
       0, & \quad x\in\Omega\backslash\Omega^\xi
      \end{array}
\right\},
\qquad y^\xi_\perp(x) := y(x)-y^\xi(x).
\end{align}
These functions have the jump discontinuities at the wave front surface
$\Gamma^\xi_\sigma$. On the front, the obvious relation holds:
$y^\xi_\perp(x(\gamma, \xi+0)) = y(x(\gamma, \xi))$.

Now, put $\partial_t v\big|_{t=T}=y^\xi_\perp$ in (\ref{dual 2})
for the dual system. Such a discontinuous perturbation produces a
wave $v^{y^\xi_\perp}$, which has a discontinuous velocity. In
particular, there is a jump of $v^{y^\xi_\perp}_t$ at the surface
$B^T_\sigma \cap \Gamma^t_\sigma$, which flies (in the inverted
time, as $t$ varies from $T$ to $T-\xi$) towards the boundary
along the rays. This jump reaches the boundary at the points
$x(\gamma,0) = \gamma \in \sigma$ at the moment $t=T-\xi$. It
produces the jump of the force, whose amplitude is also calculated
by geometric optics:
\begin{align}
\label{dualjump} \partial_\nu v^{y^\xi_\perp}(\gamma,T-\xi-0) =
\left[\frac{J(\gamma,t)J(\gamma,0)}{c(x(\gamma,t))c(x(\gamma,0))}\right]^{\frac{1}{2}}\,y(x(\gamma,\xi))\,.
\end{align}
This equality can be derived from (\ref{discont}) and the duality
relation (\ref{duality}). It is the formula (\ref{dualjump}),
which} motivates the use of the factor $\beta$ in (\ref{image}).

\subsection{Amplitude formula}
\label{sec:ampfor}

\begin{figure}[h]
\begin{center}
\includegraphics[width=0.59\columnwidth]{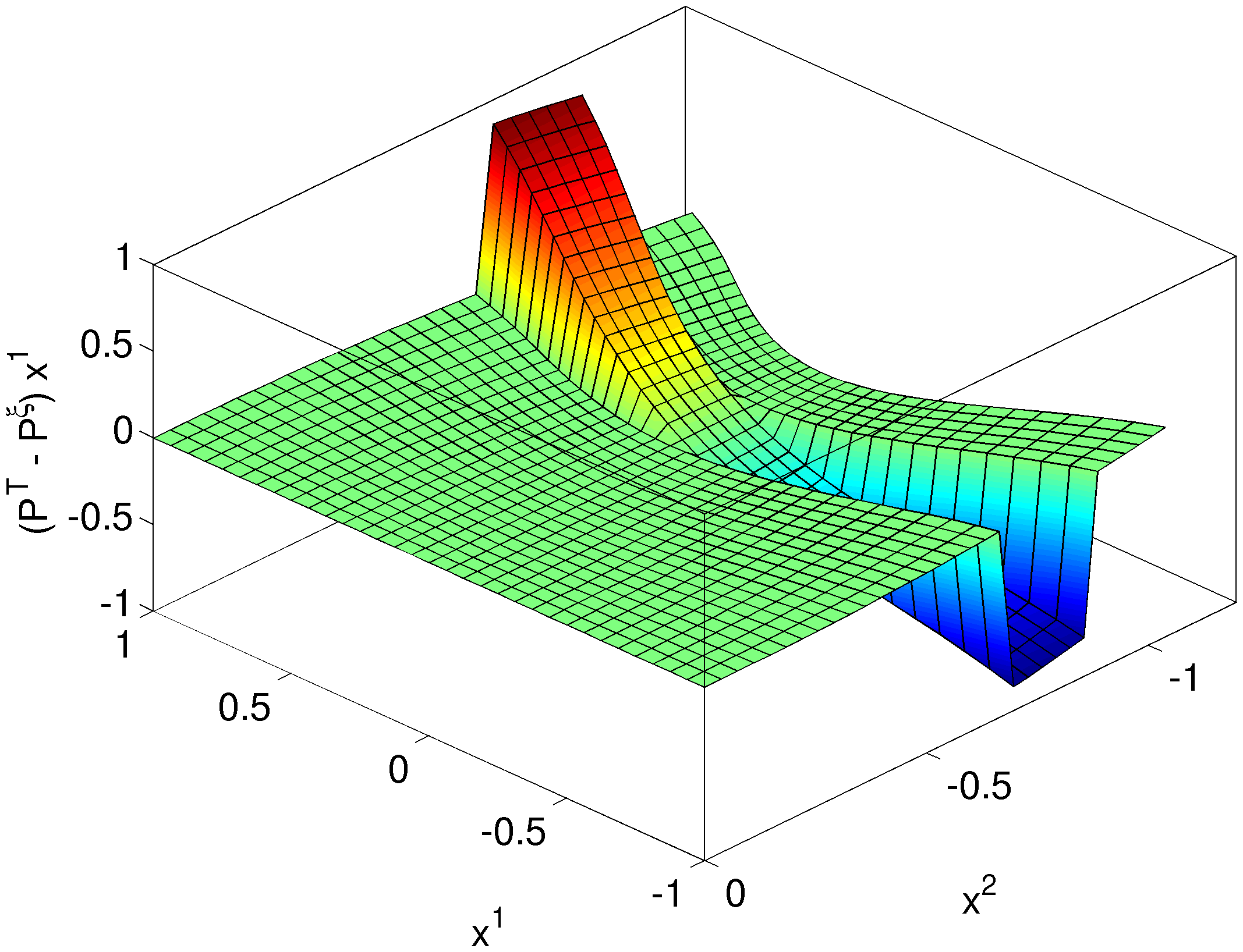}
\end{center}
\caption[]{Projection $(x^1)^\xi_\perp=(x^1)^T-(x^1)^\xi$, $T=1$, $\xi=0.75$, and $c(x)$ from Section \ref{sec:test:case_1}.}
\label{fig:proj_x1}
\end{figure}

Combining the definitions (\ref{obsop}) and (\ref{image}), the
relation (\ref{dualjump}) takes the form
\begin{align}
\nonumber \left[O^T y^\xi_\perp\right](\gamma,T-\xi-0) =
\beta(\gamma,\xi)\,y(x(\gamma, \xi)) = \tilde y(\gamma,\xi),
\qquad (\gamma, \xi) \in \Sigma^T_\sigma\,.
\end{align}
With regard to the obvious equalities
$y=y^T,\,y^\xi_\perp=y^T-y^\xi$, one can rewrite the latter
representation in the form
\begin{align}
\label{ampfor} \tilde y(\gamma,\xi) = \left[O^T(y^T - y^\xi)\right](\gamma, T-\xi-0),\qquad (\gamma, \xi) \in \Sigma^T_\sigma\,.
\end{align}
It represents the image of function as a collection of amplitudes
of the wave jumps, which pass through the medium and are observed
on the screen: see Fig.~\ref{fig:proj_x1}.

Let $\{f_\alpha^T\}$ and $\{f_\alpha^\xi\}$ be the control
families, which are complete in $\mathcal{F}^{T}_\sigma$ and
$\mathcal{F}^{T,\xi}_\sigma$ respectively, $\{u_\alpha^T\}$ and
$\{u_\alpha^\xi\}$ the corresponding wave bases in
$\mathcal{H}^{T}_\sigma$ and $\mathcal{H}^{\xi}_\sigma$. The
controllability property enables one to represent the truncated
functions by (\ref{wavexp})
\begin{align}
\label{WE} y^T \,= \,\lim_{N \to \infty}\sum \limits_{\alpha=1}^N
c^{T,\, N}_\alpha\,u^T_\alpha\,, \qquad y^\xi \,= \,\lim_{N \to
\infty}\sum \limits_{\alpha=1}^N c^{\xi,\, N}_\alpha\,u^\xi_\alpha
\end{align}
with the coefficients determined by the  linear systems (\ref{Gram
system}). Substituting the expansions (\ref{WE}) for $y^T$ and
$y^\xi$ to (\ref{ampfor}), we easily get
\begin{align}
\nonumber \tilde y(\gamma,\xi) = \left[\lim_{N \to \infty}\sum
\limits_{\alpha=1}^N\,\left(c^{T,\,N}_\alpha\,O^T u^T_\alpha -
c^{\xi,\,N}_\alpha\,O^T u^\xi_\alpha\right)\right](\gamma,
T-\xi-0)\,.
\end{align}
By (\ref{C^T=O^T W^T}), we have $O^T u^f = C^T f$ and, hence,
arrive at the representation
\begin{align}
\notag\tilde y(\gamma,\xi) = \left[\lim_{N \to \infty}\sum
\limits_{\alpha=1}^N\,\left(c^{T,\,N}_\alpha\,C^T f^T_\alpha -
c^{\xi,\,N}_\alpha\,C^T f^\xi_\alpha\right)\right](\gamma,
T-\xi-0),\qquad (\gamma, \xi) \in \Sigma^T_\sigma,
\end{align}
which is called an {\it amplitude formula} (AF). It is the
relation, which is the main computational device of the BCM.

\subsection{Visualizing waves and harmonic functions}
\label{sec:visualizing} Take a control $f \in \mathcal{F}^{T}$.
The wave $u^f$ has the part $u^f(\cdot,T)\big|_{B^T_\sigma}$,
which is a function on the tube. Such a function has an image on
the screen $\Sigma^T_\sigma$.
For this image the amplitude formula provides
 \begin{align}
\label{AF for waves}
\widetilde{\left[u^f(\cdot,T)\right]}(\gamma,\xi)\, = 
\left[\lim_{N \to \infty}\sum\limits_{\alpha=1}^N\,
\left(c^{T,\,N}_\alpha\,C^T f^T_\alpha - c^{\xi,\,N}_\alpha\,C^T f^\xi_\alpha\right)\right](\gamma,T-\xi-0),
 \end{align}
where $(\gamma, \xi) \in \Sigma^T_\sigma$ and the coefficients satisfy the linear systems of the form
(\ref{Gram system})
 \begin{align}
\label{Gram system wave} \sum \limits_{\alpha=1}^N A^s_{\beta
\alpha}c^{s,N}_\alpha\,=\,b^s_\beta, \qquad \beta=1, \dots N,
\qquad s=T,\,\xi
 \end{align}
with the matrices
\begin{align}
\label{Gram system A} A^s_{\alpha\beta} \overset{(\ref{Gram
general})}= \int\limits_{\Sigma^T_\sigma} \md t\, \md \Gamma\,
\left[\left(R^{2T}F^s_\alpha\right)(\gamma,t) -
\left(R^{2T}F^s_\alpha\right)(\gamma,2T-t)\right]\,f^s_\beta(\gamma,t)
\end{align}
and the right hand sides
\begin{align}
\label{Gram system waves b}
b^s_\beta\,=\,\left(u^f(\cdot,T),u^s_\beta\right)_{\mathcal{H}}\overset{(\ref{Gram general})}=
\int\limits_{\Sigma^T_\sigma} \md t\, \md \Gamma\,
\left[\left(R^{2T}F^s_\beta\right)(\gamma,t) - \left(R^{2T}F^s_\beta\right)(\gamma,2T-t)\right]\,f(\gamma,t)\,.
\end{align}

Now, let a function $a=a(x)$ satisfy $\Delta a=0$ in $\Omega$ and
$a\big|_{B^T_\sigma}$ be its part in the tube.
Then the amplitude formula provides
 \begin{align}
\label{AF for harmonic}
\tilde a(\gamma,\xi)\, = \,\left[\lim_{N \to \infty}\sum\limits_{\alpha=1}^N\,
\left(c^{T,\,N}_\alpha\,C^T f^T_\alpha - c^{\xi,\,N}_\alpha\,C^T f^\xi_\alpha\right)\right](\gamma,T-\xi-0),
 \end{align}
where $(\gamma, \xi) \in \Sigma^T_\sigma$ and the coefficients satisfy the linear systems (\ref{Gram system wave})
with the same matrices $A^s_{\alpha \beta}$ given by (\ref{Gram system A}) but the right hand sides of the form
\begin{align}
\label{Gram system harmonic b} b^s_\beta\,=\,\left(a,
u^s_\beta\right)_{\mathcal{H}}\overset{(\ref{(a, u^f)})} =\,
\int\limits_{\Sigma^T} \md t\, \md
\Gamma\,(T-t)\,\left[a(\gamma)(R^Tf^s_\beta)(\gamma,t)-\partial_\nu
a(\gamma)\,f^s_\beta(\gamma,t)\right]\,.
\end{align}

The external observer applies the controls $f_\alpha$ on the
boundary part $\sigma$ and measures the response $R^{2T}f_\alpha$.
As the representations (\ref{AF for waves})--(\ref{Gram system
harmonic b}) show, such measurements suffice to recover the images
of waves and harmonic functions on the screen $\Sigma^T_\sigma$.
In the BCM such an option is referred to as a {\it visualization}.

\section{BCM reconstruction}\label{sec:recon}

\subsection{How BCM works}\label{sec:how_bcm_works}

To recover $c$ in the tube $B^T_\sigma$ we summarize the previous mathematical considerations to show how the BCM actually works.

A central object of the BCM is the {\it dual} dynamic system (\ref{dual 1})--(\ref{dual 3}).
It is the acoustic equation with the same speed of sound $c(x)$ as in normal dynamic system (\ref{acoustic 1})--(\ref{acoustic 3})
but it is considered in inverted time,
with zero Dirichlet boundary conditions on $\Gamma$, with zero initial value of a solution $v^y|_{t=T} = 0$ at time instance $t = T$, 
and with some given initial value of time derivative of a solution $\partial_t v^y|_{t=T} = y(x)$.
If the initial data $y(x)$ have a jump discontinuity then well known {\it exact}
result (\ref{dualjump}) of geometric optics can be used to relate an amplitude of the jump in normal derivative of the solution 
$\partial_\nu v^y|_{\Sigma^T}$ with the amplitude of the jump in $y(x)$.
Thus, using projections (\ref{projector}) of harmonic functions $(1, x^1, x^2, \dots, x^n)$ on subdomains $\Omega^\xi$ filled by
waves at different time instances $\xi$ as the discontinuous initial data $y(x)$
we may, by means of measurements of the normal derivative $\partial_\nu v^y|_{\Sigma^T}$ at 
points on the domain boundary $\Gamma$, extract information about values of the harmonic functions
inside the domain $\Omega^T$ at points connected by the geodesic lines (acoustic rays) with the observation points on $\Gamma$.
In this way the {\it semigeodesic} coordinates and hence a speed of sound inside $\Omega^T$ can be recovered.
The questions are: how to prepare the required discontinuous initial data for the dual dynamic system, and how to calculate
the normal derivative of its solution at different times on the boundary $\Gamma$
as only normal dynamic system is at out disposal.

For the first task we make use of the property of {\it approximate controllability} of the dynamic system
which means that any square-integrable function $y(x)$ can be represented with any prescribed accuracy as a superposition 
of wave solutions $u^{f_\alpha}(\cdot,T)$ produced by some (full) set of boundary controls $f_\alpha$.
To find the expansion coefficients for the harmonic functions the linear systems (\ref{Gram system})
have to be solved whose matrix consists of scalar products
of the wave solutions and the right hand sides consist of scalar products of the wave solutions 
and the functions being expanded. The remarkable fact is that all these {\it internal} scalar products
can be calculated {\it exactly} from some {\it external} data available on the domain boundary,
see the expressions (\ref{Gram system A}) and (\ref{Gram system harmonic b}).
In the result, by means of the controls $f_\alpha$ acting on the boundary we create
the required projections of harmonic functions on subdomains $\Omega^\xi$ filled by waves at different time instances $\xi$.

For the second task we make use of another remarkable fact.
If we take a wave solution produced by a boundary control $f$ as initial value for time derivative
$\partial_t v^y |_{t=T} = y = u^f(\cdot,T)$ in the dual dynamic system then normal derivative of the solution
$\partial_\nu v^y|_{\Sigma^T}$ can be obtained measuring the response of the {\it normal} dynamic system on some
{\it double} boundary control $F$ related to the control $f$ in a simple manner,
see (\ref{C^T=O^T W^T}) and (\ref{C^T via R^2T}).

Finally, combining together all tricks we come to so called {\it amplitude formula} (\ref{AF for harmonic})
which allows us to calculate the images of harmonic functions in
a tube covered by direct rays from measurements of normal derivative
of wave solutions produced by some (full) set of boundary controls $f_\alpha$ and their double counterparts $F_\alpha$.

For a point $x=(x^1, \dots , x^n) \in {\bf R}^n$, we consider its
$k$-th component as a function of $x$ and write $x^k(x)$. These
coordinate functions are harmonic: $\Delta x^k(\cdot) = 0$. By
$1(\cdot)$ we denote the function equal to $1$ identically; it is
also harmonic. All these functions have the images, which can be
recovered by the amplitude formula. The following procedure just
exploits such an option.
\smallskip

\noindent{\bf Step 1.}\,\, Fix a $\xi<T$. Choose the families of
controls $\{f_\alpha^T\}$ and $\{f_\alpha^\xi\}$, which are
complete in $\mathcal{F}^{T}_\sigma$ and
$\mathcal{F}^{T,\xi}_\sigma$ respectively. Applying (\ref{AF for
harmonic}) and (\ref{Gram system harmonic b}), find the images
 $$
\tilde x^k(\gamma, \xi)=\beta(\gamma, \xi)x^k(x(\gamma,\xi)),
 \,\,\,k=1, \dots, n; \quad
\,\tilde 1(\gamma, \xi)=\beta(\gamma, \xi)\,.
 $$
\noindent{\bf Step 2.}\,\,Varying $\xi$, find the images $\tilde
x^k$ on the whole $\Sigma^T_\sigma$. Thereby, the connection
between the ray and Cartesian coordinates is revealed and given by
the correspondence:
$$
\Sigma^T_\sigma \ni (\gamma, \xi)\mapsto x(\gamma,
\xi)=\left(\frac{\tilde x^1(\gamma, \xi)}{\tilde 1(\gamma, \xi)},
\dots, \frac{\tilde x^n(\gamma, \xi)}{\tilde 1(\gamma,
\xi)}\right) \in B^T_\sigma\,.
 $$
So, the external observer recovers the tube in $\Omega$.
\smallskip

\noindent{\bf Step 3.}\,\ Differentiation with respect to $\xi$
corresponds to differentiation along the ray in the tube that
implies
\begin{equation}
\label{speed} c(x(\gamma, \xi)) =
\left\{\sum_{k=1}^n\left[\frac{\partial{x^k\left(x(\gamma,\xi)\right)}}
{\partial{\xi}}\right]^2\right\}^\frac{1}{2},\qquad (\gamma, \xi)
\in \Sigma^T_\sigma\,.
\end{equation}
The pairs $\{x(\gamma,\xi),c(x(\gamma,\xi))\}$ for all
$(\gamma,\xi) \in \Sigma^T_\sigma$ constitute the graph of $c$ in
$B^T_\sigma$. {\it Thus, $c|_{B^T_\sigma}$ is determined}.

\noindent{\bf Remark.}\,\,\,Our procedure recovers the speed of
sound in the ray tube, excluding the cut locus. The controls
$f_\alpha$ prospecting the tube are supported on its bottom
$\sigma$ but the response $R^{2T}f_\alpha$ has to be measured on a
wider part $\Omega^T_\sigma \cap \Gamma$ of the boundary. There is
a version which recovers $c|_{B^T_\sigma}$ via $R^{2T}f_\alpha$
given on $\sigma$ only: see \citep{belivsem-2015}, section 4.1.
However, it is problematic for numerical implementation.

\subsection{Choice of controls}\label{sec:basis}

The BCM uses the boundary controls $f_\alpha$, which are
continuous, vanish at $t=0$, and constitute a complete system in
the proper space. We construct a basis of boundary controls by
direct product of spatial and temporal bases, $f_\alpha(\gamma,t)
= \phi_l(\gamma)\,\psi_m(t)$, $\alpha=l + m N_\gamma$, where
$l=\range{0}{N_\gamma-1}$, $m=\range{0}{N_t-1}$, and the basis
dimension is $N = N_\gamma N_t$.

In case of semiplane we can keep under control only a part of the
boundary and thus have to use {\it localized} spatial basis
functions. The simplest and good choice is conventional
trigonometric basis localized to interval [-1, 1] by an
exponential cutoff multiplier \citep{belivsem-2015}. The commonly
used in FEA tent functions are another convenient choice which has
several advantages: the simplicity of hardware implementation and
the same spatial scale of all basis elements. Moreover, the tent
functions are also very suitable for construction of the temporal
basis since all basis functions (and corresponding solutions) can
be obtained from the first one just by delays in time. However,
the tent functions have discontinuous derivative and the rate of
convergence of conventional numerical methods for generation of
synthetic inverse data is low.

\begin{figure*}
\begin{center}
\includegraphics[width=0.7\columnwidth]{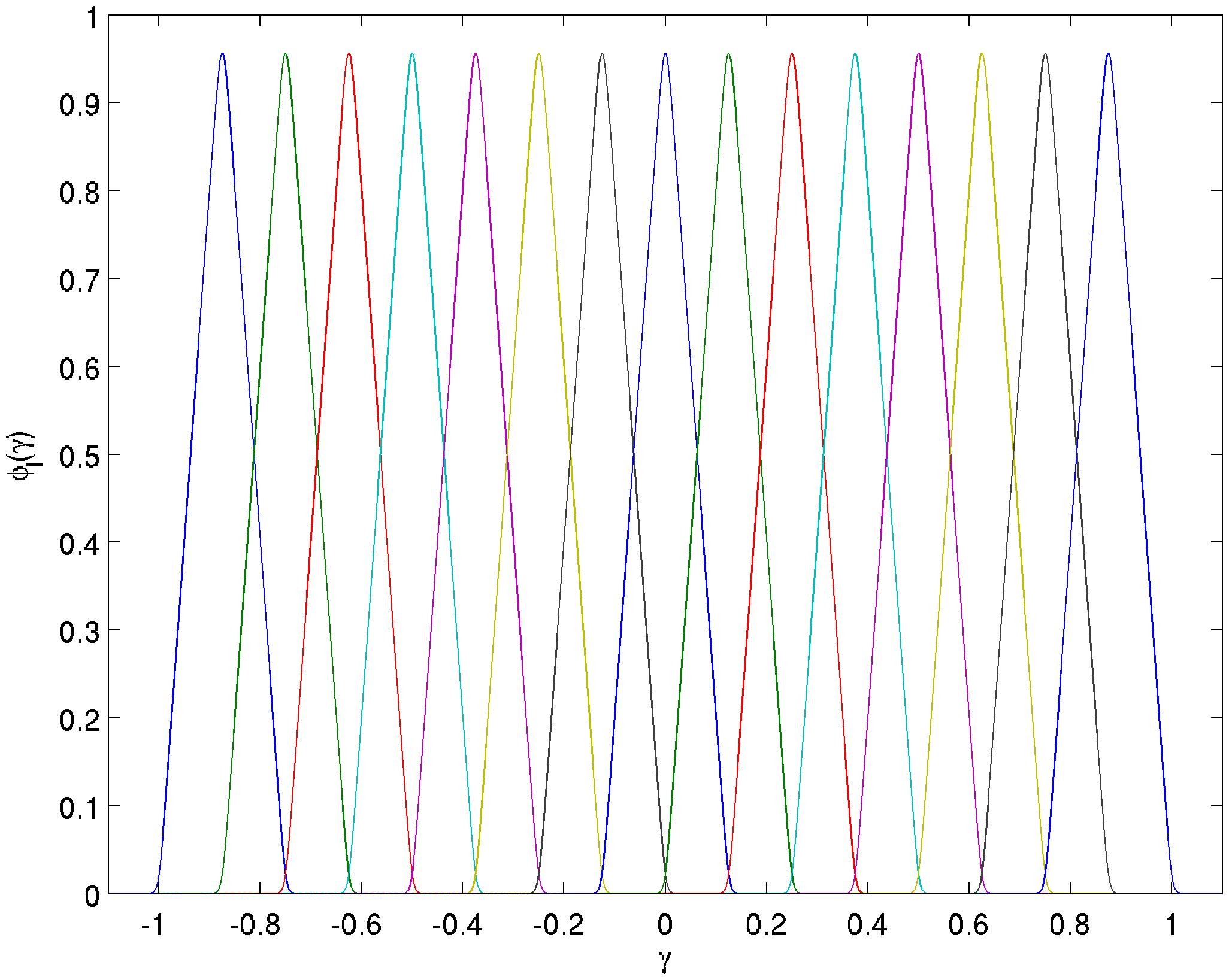}\vspace{1cm}
\includegraphics[width=0.7\columnwidth]{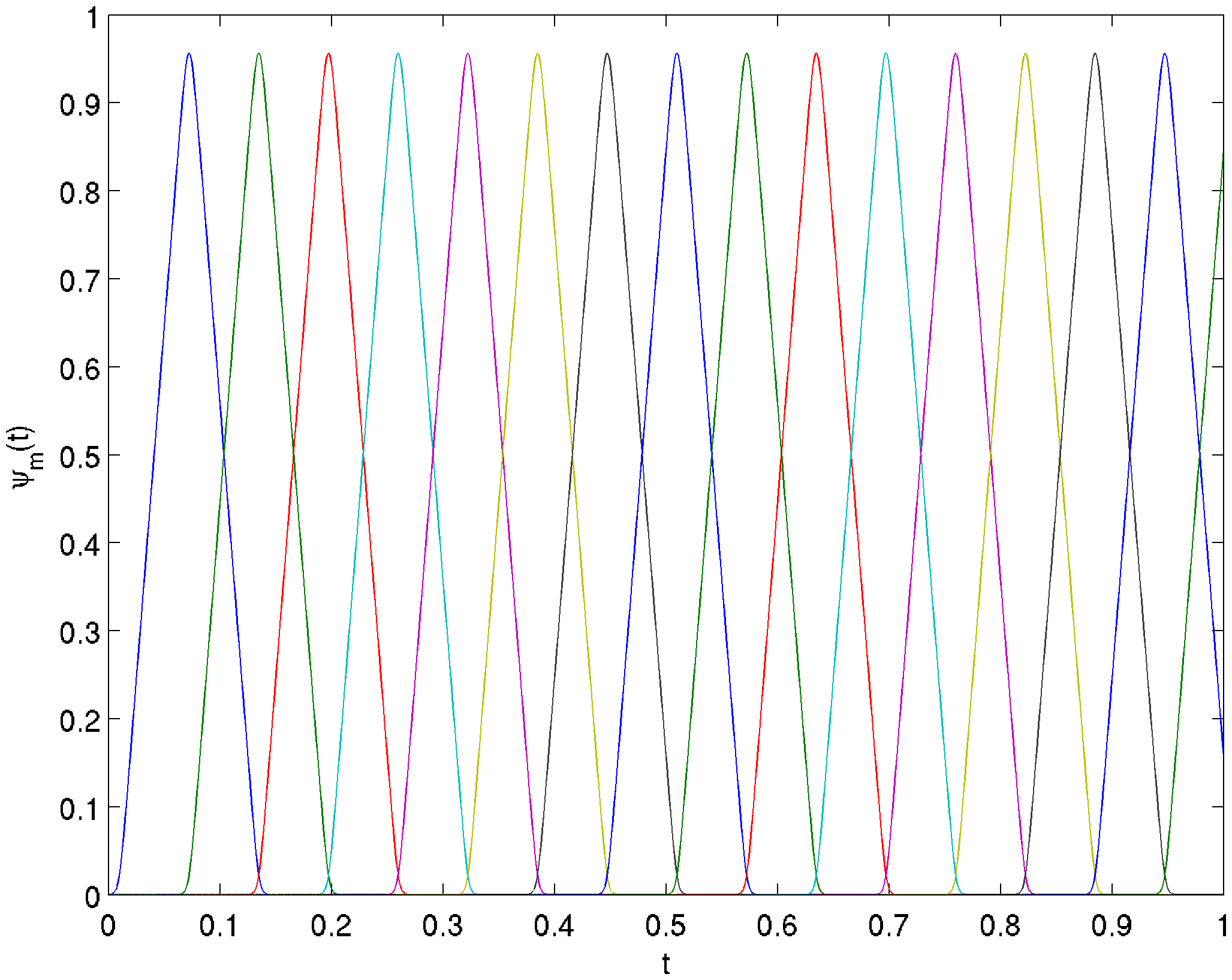}
\end{center}
\caption{Basis of boundary controls: 15 spatial functions $\phi_l(\gamma)$ on interval $[-1,1]$ (top),
and 16 temporal functions $\psi_m(t)$ on interval $[0,T]$ with $T = 1$ (bottom). In both cases $d = \Delta/32$.}
\label{fig:basis}
\end{figure*}

To overcome this difficulty we have used a {\it smooth tent-like}
function,
\begin{equation}
\nonumber \theta(z) = \frac{d}{\Delta}\, \ln\left[ \frac {
\cosh\left(\frac{2\Delta-z}{2 d}\right) \cosh\left(\frac{z}{2
d}\right) } { \cosh^2\left(\frac{\Delta-z}{2 d}\right) } \right] /
\left[1 - \exp\left(-\frac{\Delta}{d}\right)\right],
\end{equation}
where $z$ is an independent variable ($\gamma$ or $t$), $2\Delta$
is the triangle base, and $d$ is a smoothing parameter (when
$d\rightarrow 0$ function $\theta(z)$ gets a triangular shape).
The spatial functions on interval $\gamma\in[a, b]$ are
$\phi_l(\gamma) = \theta(\gamma - a - l\Delta)$, where $\Delta =
(b - a) / (N_\gamma + 1)$. The temporal functions on interval
$t\in[0, T]$ are $\psi_m(t) = \theta(t - \delta - m\Delta)$, where
$\Delta = T / N_t$, and $\delta$ is a small offset to ensure a
negligible value of $\theta(0)$. Such simple and translationally
invariant basis shown in Fig.~\ref{fig:basis} considerably reduces
computational resources needed for BCM reconstruction.

\subsection{Regularization}

\label{sec:regularization}

In BCM we expand discontinuous projections $a^\xi=P^\xi a$ over smooth
wave solutions $u^\xi_\alpha(\cdot,T)$ and thus have to observe
the {\it Gibbs oscillations}. The basis functions have a finite
resolution of order of spatial and temporal scales of the
tent-like functions. All scales below the minimum ones are
unreachable, therefore we {\it average} a result of expansion
$g(\gamma, t)$ over that minimum scales by convolution with some
kernel $K(\gamma, \xi)$,
\begin{align}
\label{convol} \langle g\rangle(\gamma,t) :=
\int\limits_{-\infty}^{+\infty}\md
t'\int\limits_{-\infty}^{+\infty}\md\gamma' K(\gamma - \gamma', t
- t')\,g(\gamma', t').
\end{align}
In our implementation the kernel $K(\gamma - \gamma', t - t')$ is
a product of conventional {\it Gaussian kernels} both for spatial
and temporal variables. Such procedure efficiently removes the
Gibbs oscillations and thus {\it de facto} accelerates convergence
of the expansions. The values of standard deviations in Gaussian
kernels should match the minimum spatial and temporal scales of
the boundary controls to efficiently smooth out the Gibbs
oscillations.

Due to unavoidable errors in matrix elements and right hand sides
of linear systems, the expansion coefficients $c^{\xi,\,N}_\alpha$ also
contain errors amplified by ill-conditioned matrix $A$. In some
cases we have to use Tikhonov (or other) regularization to reduce
additional fake oscillations caused by errors in expansion
coefficients. The value of the regularization parameter is
selected to satisfy a desired tolerance for residual of the linear
systems.

\section{Numerical testing}

\label{sec:test}

We have performed detailed tests of quality of the BCM
reconstruction on semiplane (with trigonometric spatial basis) in
\citep{belivsem-2015}. The method have demonstrated a good accuracy
of reconstruction of $c(x)$ (of order of several percents in most
part of the recovered domain) in cases of regular field of
acoustic rays.

The goal of the following tests is to try the method with the
tent-like spatial basis functions and in case of irregular
realistic field of acoustic rays (Marmousi model).

For each delayed control $f^\xi_\alpha(\gamma,t)$ and its double
version $F^\xi_\alpha(\gamma,t)$, we have to find a solution of
the direct problem (\ref{acoustic 1})-(\ref{acoustic 3}) and to measure the system's
reaction at the boundary $\Gamma$. For this purpose, we have
implemented an universal semidiscrete central-upwind third order
accurate numerical scheme with WENO reconstruction suggested in
\citep{kurganov2001}.

The reaction data are then used in the procedure described in
Section \ref{sec:how_bcm_works} to reconstruct a speed of sound
$c(x)$. The procedure is simple and efficient, essentially it
involves a calculation of quadratures (double sums) for scalar
products and a solution of the resulting linear systems by
standard LAPACK routines.

\subsection{Case 1}

\label{sec:test:case_1}

We take a speed of sound $c(x) = \rho(x)^{-\frac{1}{2}}$ produced
by the density of medium,
\begin{align}
\nonumber
\rho(x^1, x^2) = 1 + a\,g_1(x^1)\,g_2(x^2),\\
\nonumber \quad g_k(x^k) = \exp{\left[-\frac{\left(x^k - \bar
x^k\right)^2}{2\Delta_k^2}\right]},
\end{align}
where $a = 1$, $\bar x^1 = 0$, $\bar x^2 = -0.5$, $\Delta_1 =
0.5$, $\Delta_2 = 0.5$.

The boundary control is applied on a part of the boundary
$\gamma\in(-1, 1)$ with probing time $T = 1$. The field of
acoustic rays shown in Fig.~\ref{fig:sgc} is regular in the
prospected domain. The basis of controls is composed from
$N_\gamma = 15$ tent-like spatial functions and $N_t = 16$
tent-like temporal functions. The dimension of each basis is
selected to obtain a desired spatial resolution and accuracy of
reconstruction.

\begin{figure*}
\begin{center}
\includegraphics[width=0.75\columnwidth]{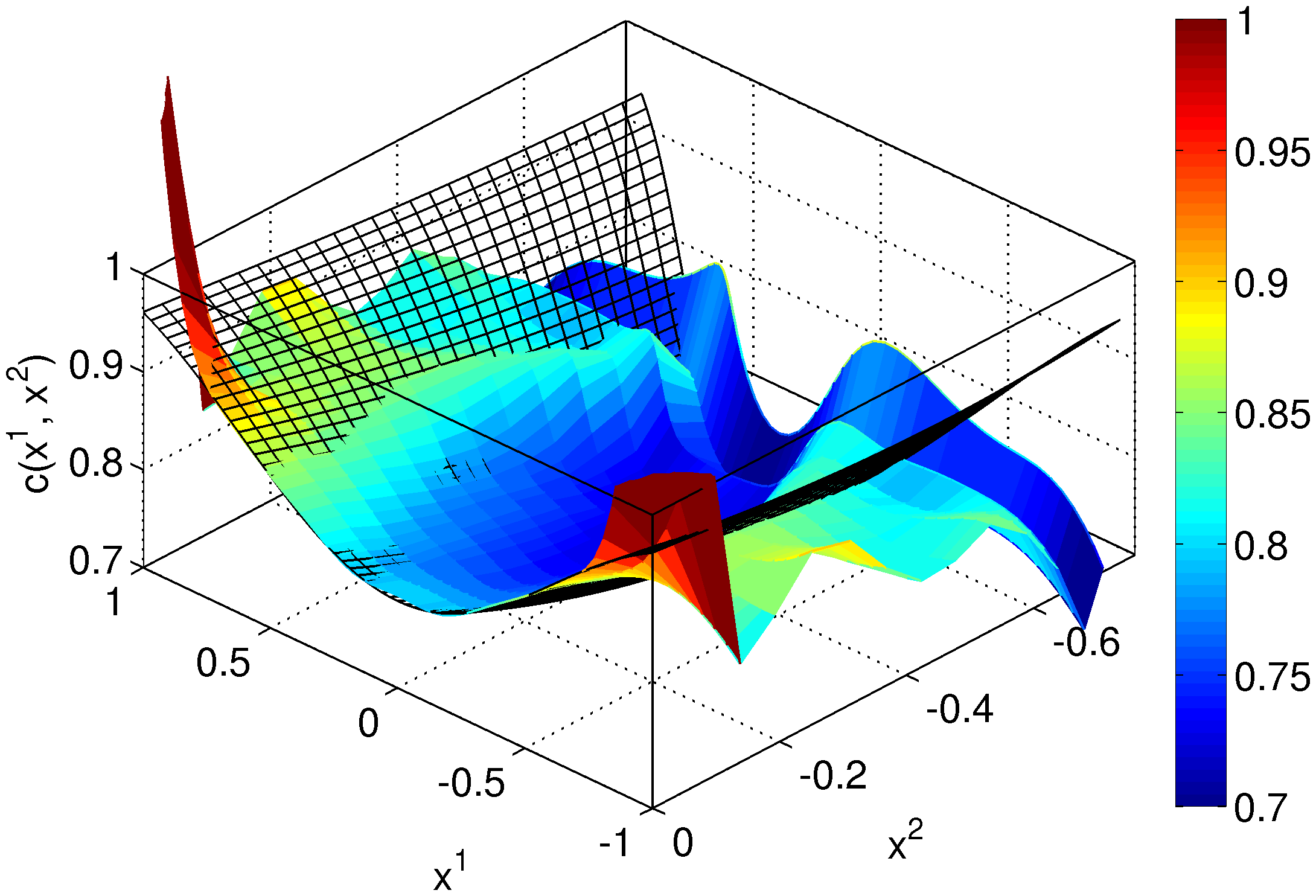}\vspace{1cm}
\includegraphics[width=0.75\columnwidth]{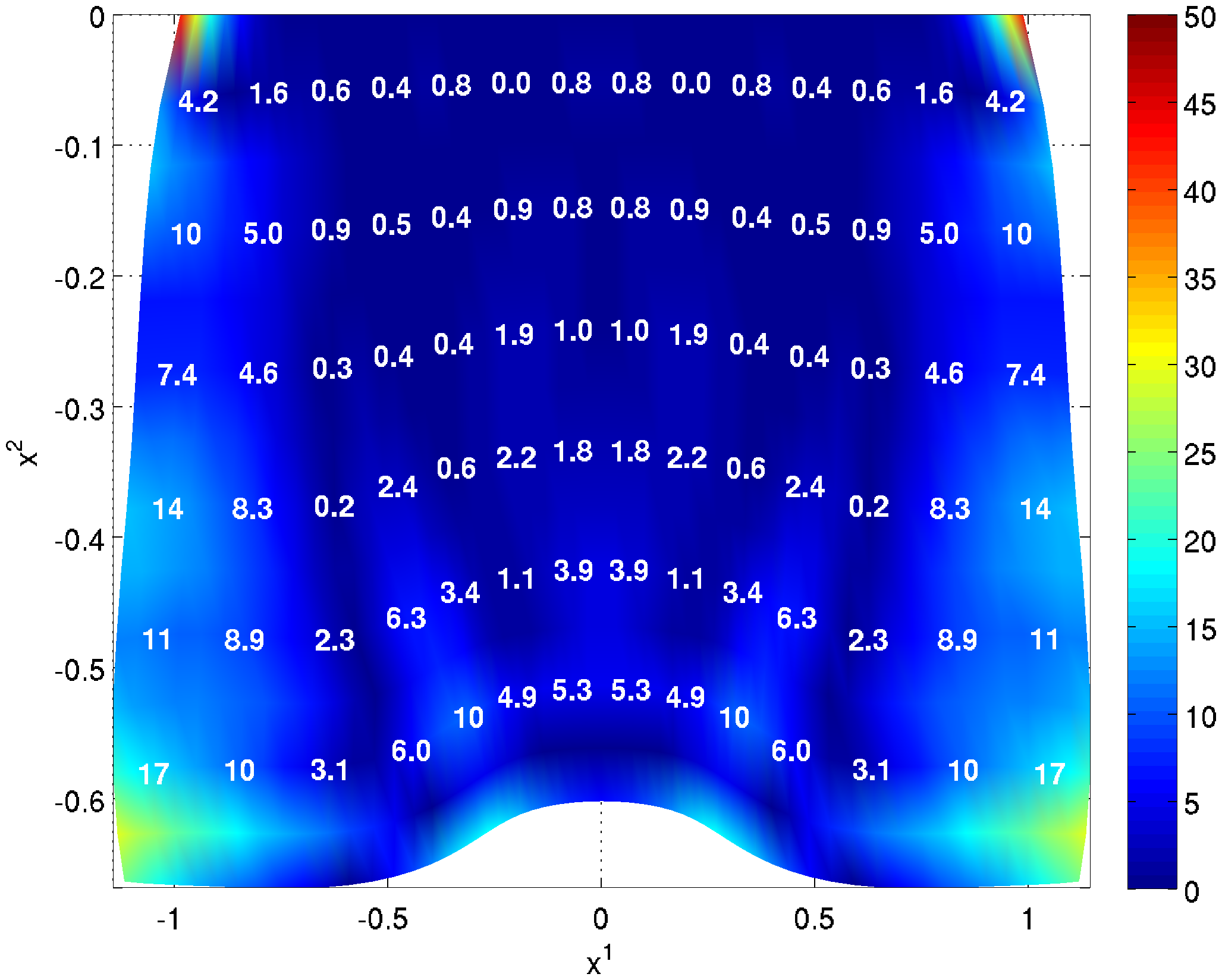}
\end{center}
\caption[]{Test 1. Top - exact (mesh) and recovered (surface) values of speed of sound $c(x)$,
bottom - map of relative errors (in percents) of the recovered values of $c(x)$ in cartesian coordinates.}
\label{fig:case1}
\end{figure*}

The exact and recovered values of speed of sound $c(x)$ are shown
in the top panel of Fig.~\ref{fig:case1} while the relative
errors of reconstruction are shown in the bottom panel of the
figure. The parameter of Tikhonov regularization for all linear
systems is $\varepsilon = 1{\cdot}10^{-6}$, and standard
deviations of Gaussian kernels in (\ref{convol}) for $(\gamma, t)$
are $\sigma_\gamma = 0.125$ and $\sigma_t = 0.0625$. We observe
that in most part of the domain covered by direct rays from the
active boundary interval $\sigma$ the relative error does not
exceed a few percents. However, as expected, the reconstruction
error grows towards the lateral borders of the recovered domain
and for large values of $\xi\approx T$. This is the effect of
degraded quality of harmonic functions projections produced by
controls acting from the boundary.

\subsection{Case 2}

\label{sec:test:case_2}

\begin{figure*}
\begin{center}
\includegraphics[width=\columnwidth]{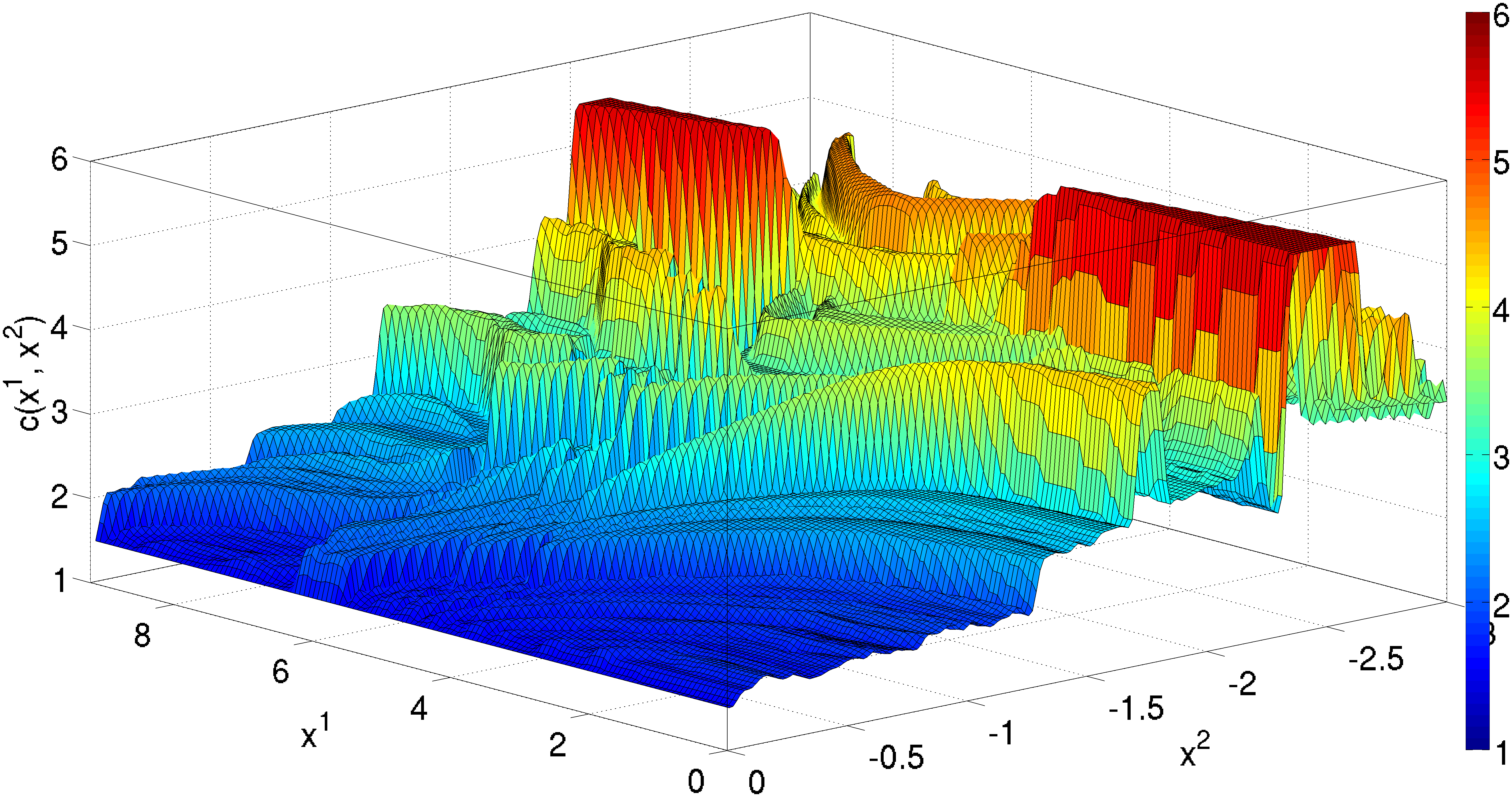}\vspace{1cm}
\includegraphics[width=\columnwidth]{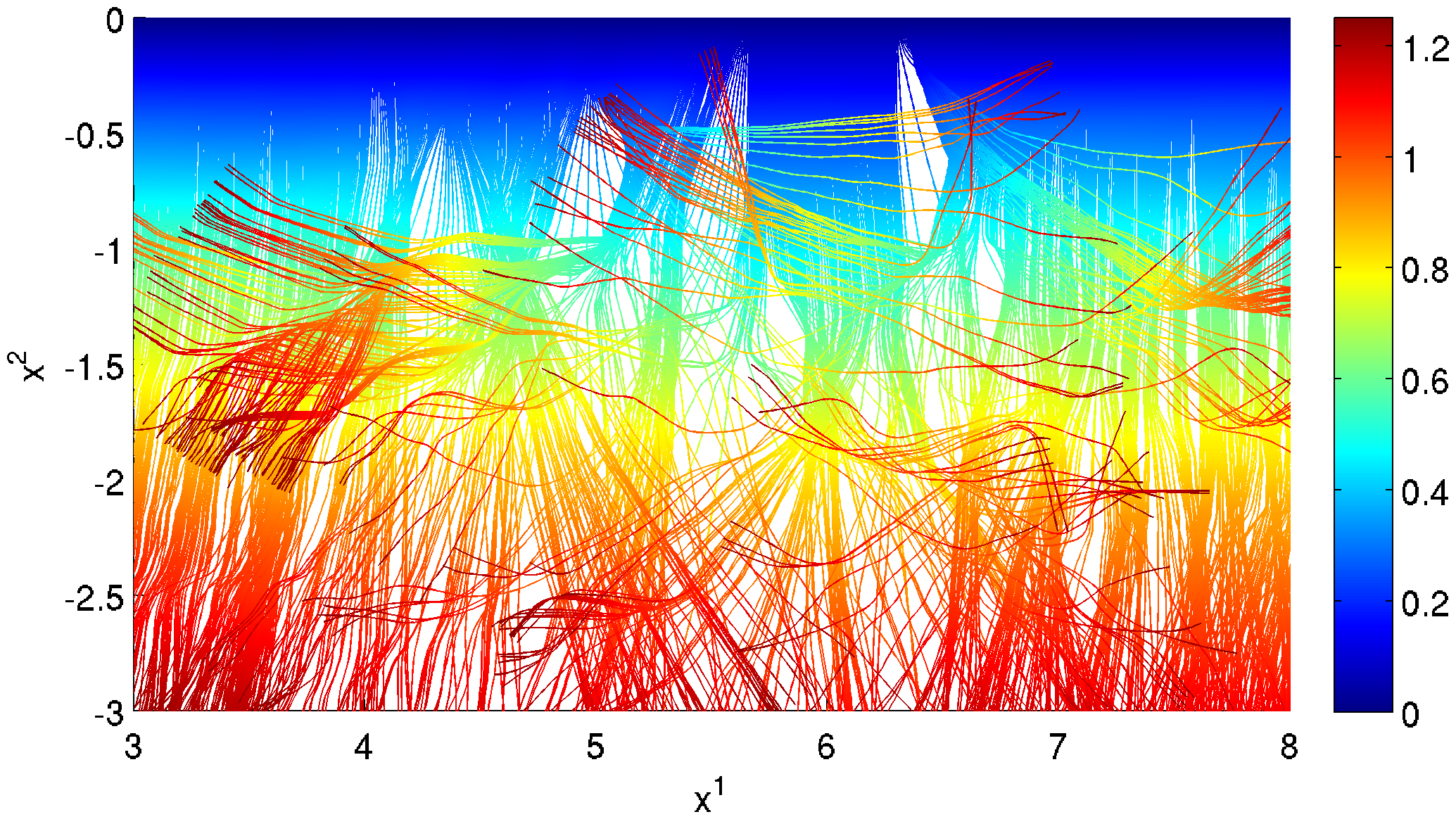}
\end{center}
\caption[]{Test 2. Top - speed of sound in the smoothed Marmousi model, 
bottom - field of acoustic rays $x(\gamma, \xi)$ emanating orthogonally from $\sigma$ with probing time $T = 1.25$ s.}
\label{fig:case2}
\end{figure*}

For the second test we have selected a speed of sound from famous
Marmousi model which has extremely complicated and irregular field
of acoustic rays, see the bottom panel of Fig.~\ref{fig:case2}. 
The original discontinuous and spiky Marmousi speed data were slightly
smoothed out by convolution with a short scale bump kernel for
better accuracy of numerical simulations of inverse data. We also
smoothly extended the original speed of sound to a larger spatial
domain to test the BCM reconstruction on the whole domain with
Marmousi data. The resulting speed of sound (without the
extension) is shown in the top panel of Fig.~\ref{fig:case2}.
The Marmousi density data were not used in our one-parameter acoustic equation.

The boundary control is done on a part of the boundary
$\gamma\in(0, 9.2)$ km with probing time $T = 1.25$ s which is
enough to prospect the domain up to $3$ km below the boundary. The
basis of controls is composed from $N_\gamma = 31$ tent-like
spatial functions and $N_t = 32$ tent-like temporal functions. The
standard deviations of Gaussian kernels are $\sigma_\gamma =
0.2875$ km and $\sigma_t = 0.0390625$ s.

\begin{figure*}
\begin{center}
\includegraphics[width=\columnwidth]{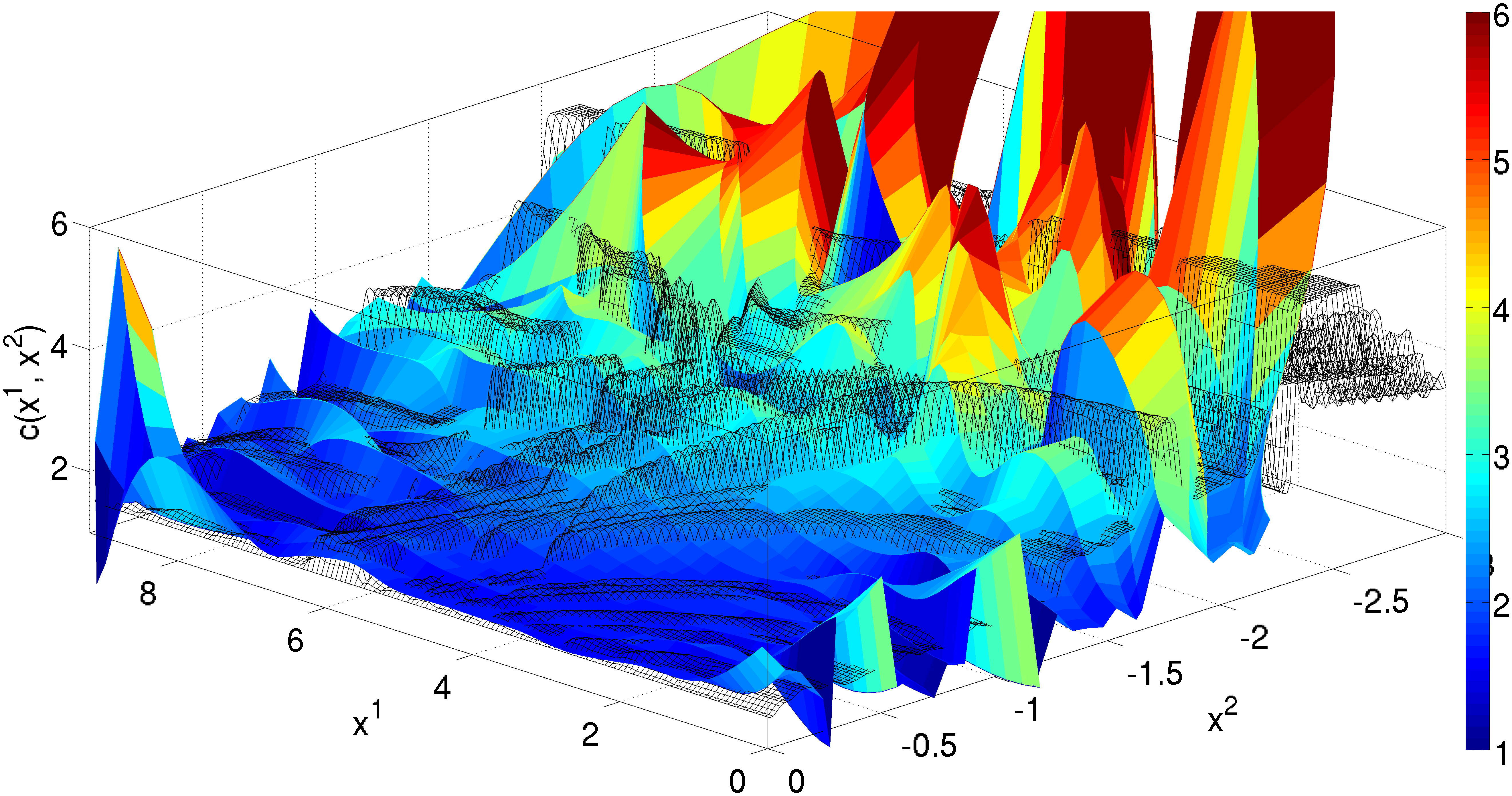}\vspace{1cm}
\includegraphics[width=\columnwidth]{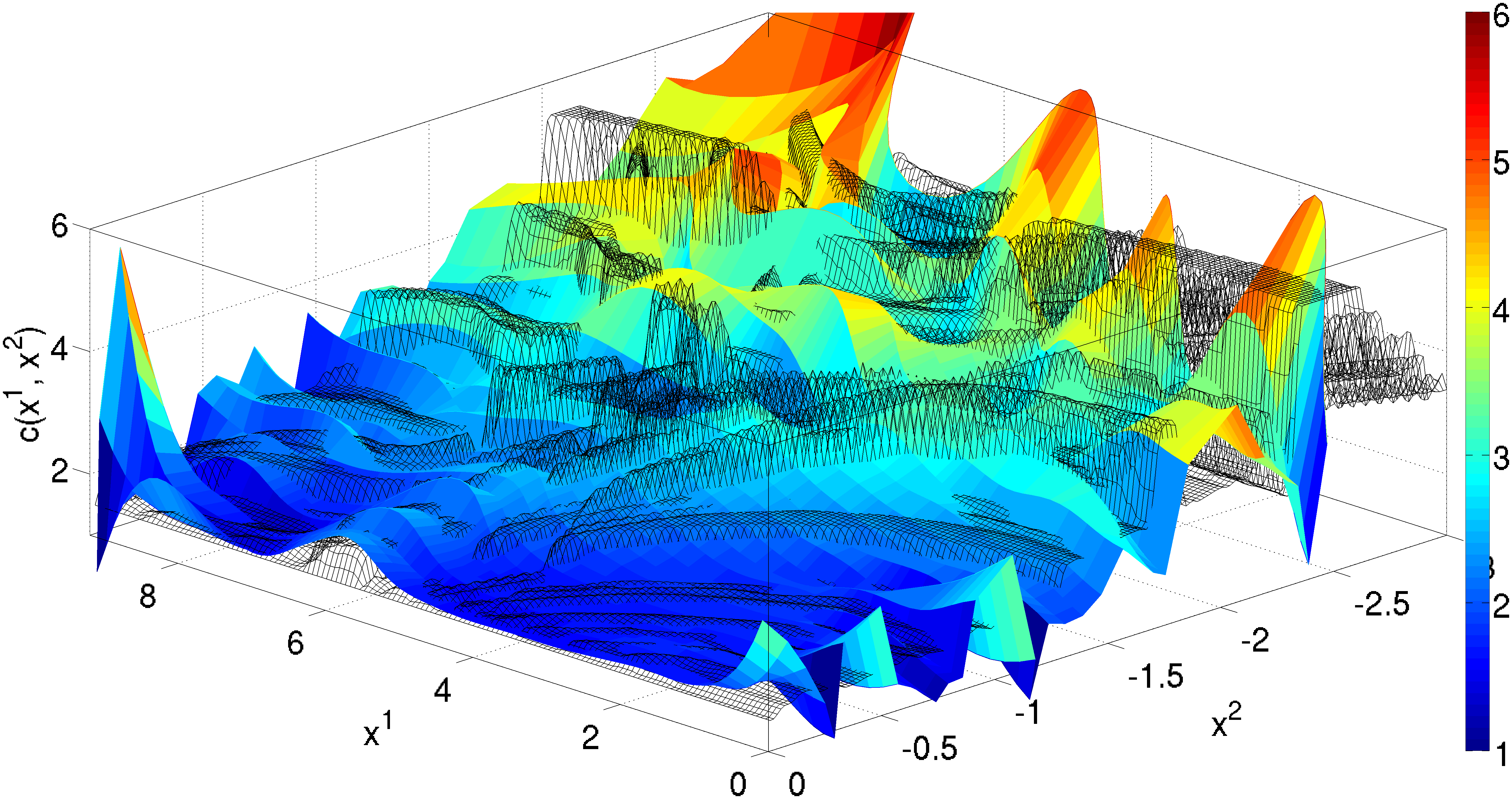}
\end{center}
\caption[]{Test 2. The speed of sound in Marmousi model: top - normal reconstruction (${\epsilon=1\cdot10^{-5}}$),
bottom - pseudo-reconstruction (${\epsilon=5\cdot10^{-7}}$).
The exact values of $c(x)$ in both panels are shown by mesh.}
\label{fig:case2:rec}
\end{figure*}

\begin{figure*}
\begin{center}
\includegraphics[width=\columnwidth]{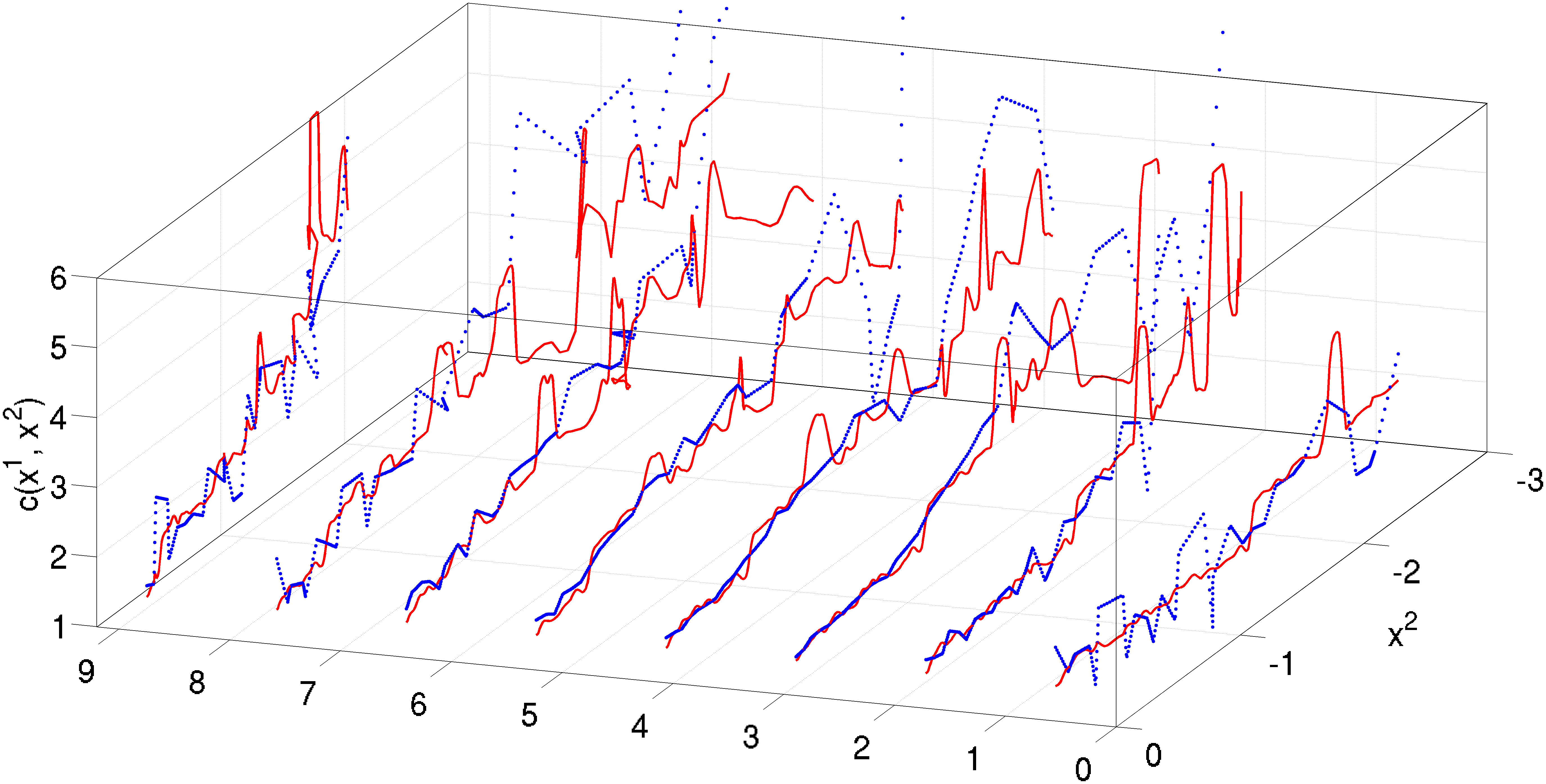}\vspace{1cm}
\includegraphics[width=\columnwidth]{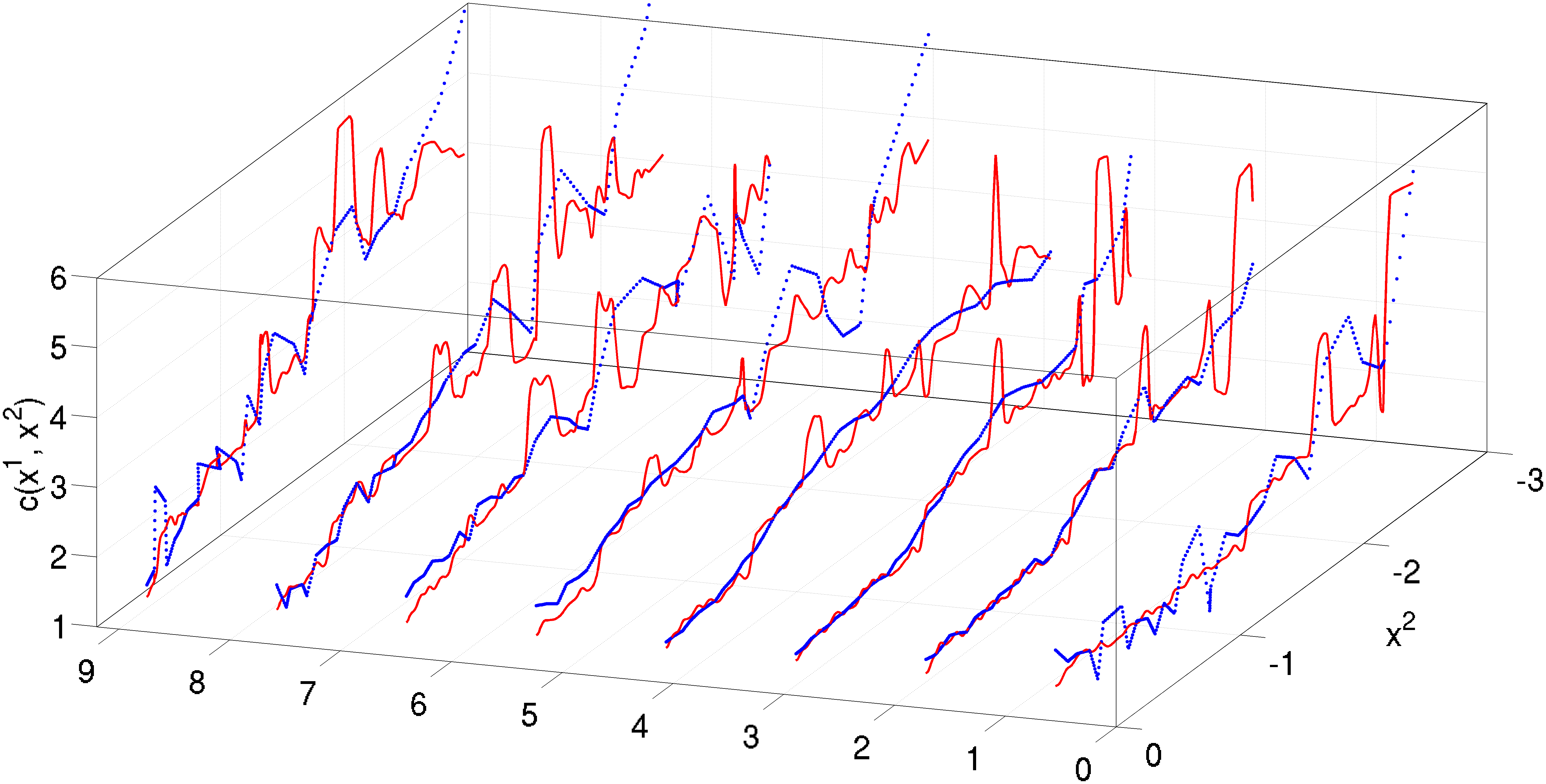}
\end{center}
\caption[]{Test 2. The profiles of speed of sound in Marmousi model:
top - normal reconstruction (${\epsilon=1\cdot10^{-5}}$),
bottom - pseudo-reconstruction (${\epsilon=5\cdot10^{-7}}$).
The exact values are shown by solid lines, the recovered values are shown by dots.}
\label{fig:case2:prof}
\end{figure*}

It is obvious that resolution of the given basis is not enough to
reconstruct sharp features of Marmousi model. As expected, in such
case the BCM procedure recovers an {\it averaged} profile of
the speed of sound, see the top panels of Fig.~\ref{fig:case2:rec} and Fig.~\ref{fig:case2:prof}.
To understand the reason of large
reconstruction errors at depths with $x^2 < - 2$ km we also
performed so called {\it pseudo-reconstruction}, see the bottom
panels of Fig.~\ref{fig:case2:rec} and Fig.~\ref{fig:case2:prof}.
It means the {\it same} reconstruction procedure of Section \ref{sec:how_bcm_works} but with
all scalar products (in linear systems) evaluated by more accurate
{\it internal} quadratures (\ref{inner}) instead of {\it external}
ones (\ref{outer}). We see that pseudo-reconstruction has a
similar accuracy at small to moderate depths but it is much more
accurate at large depths. This can be explained as follows. The
condition number of matrices $A^\xi$ (\ref{Gram system A}) grows as $\xi^4$, and it
reaches values of order $10^5 - 10^6$ at $\xi\sim T$. The errors
of the expansion coefficients $c^{\xi,\,N}$ are the errors of the right
hand sides $b^\xi$ multiplied on the condition number of $A^\xi$.
For large depths of recovering the accuracy of external scalar
products is not enough to provide accurate values of expansion
coefficients and (after all reconstruction steps) speed of
sound.

\begin{figure*}
\begin{center}
\includegraphics[width=\columnwidth]{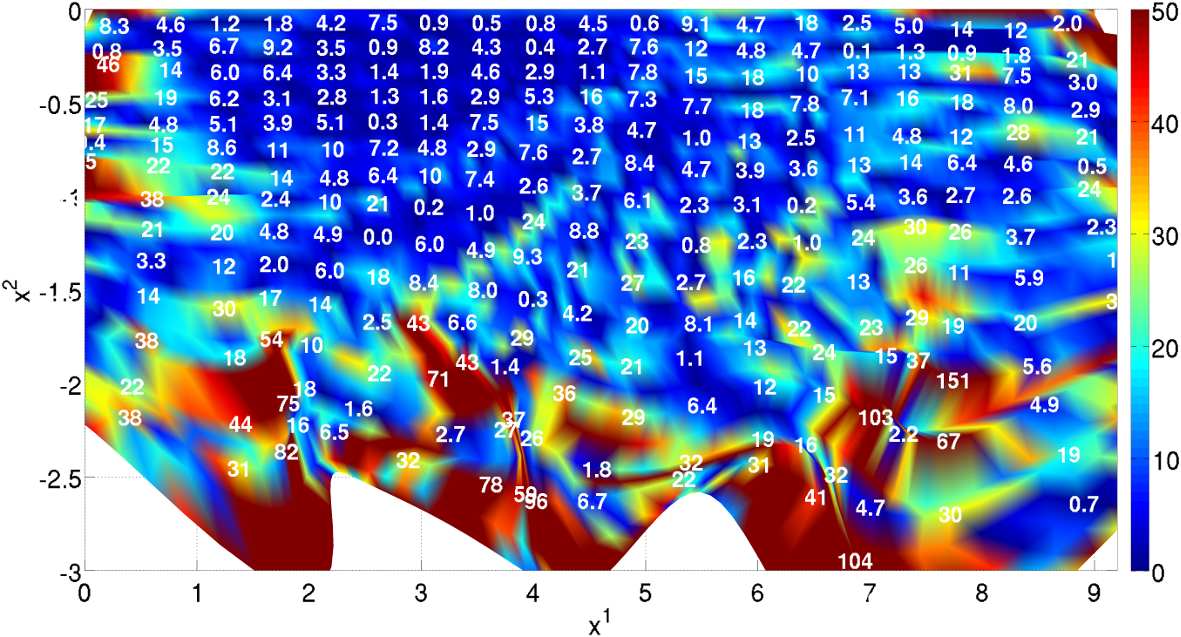}\vspace{1cm}
\includegraphics[width=\columnwidth]{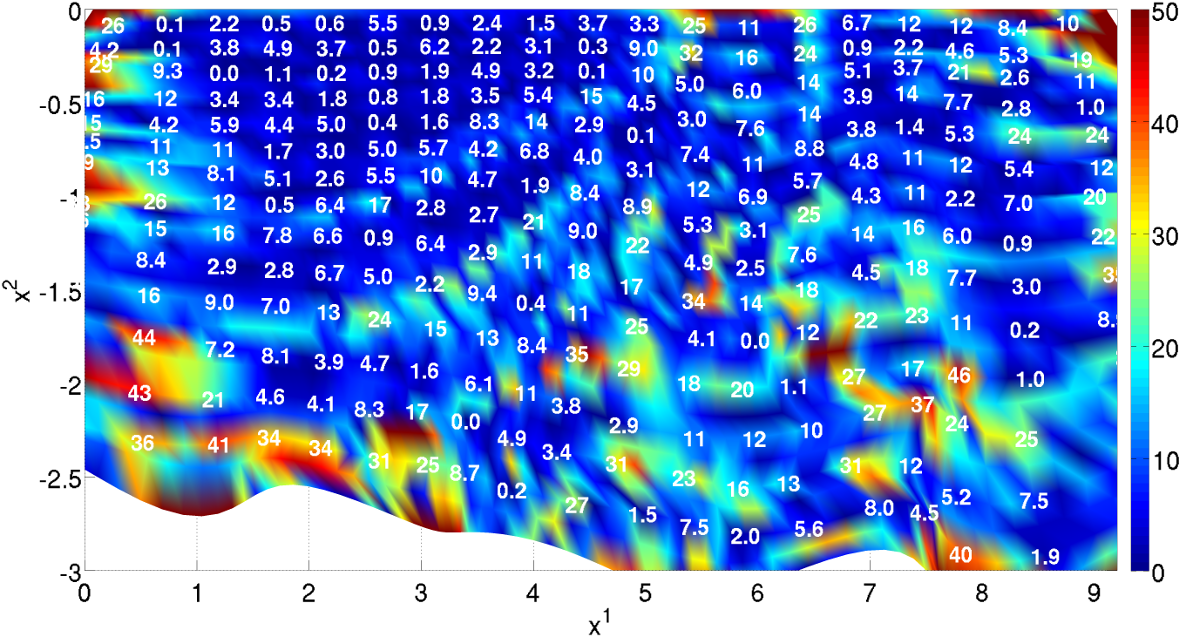}
\end{center}
\caption[]{Test 2. The map of relative errors of speed of sound in percents:
top - normal reconstruction (${\epsilon = 1\cdot10^{-5}}$),
bottom - pseudo-reconstruction (${\epsilon = 5\cdot10^{-7}}$).}
\label{fig:case2:relerr}
\end{figure*}

The relative errors of the recovered speed of sound are shown in
Fig.~\ref{fig:case2:relerr}. We see that in the most part of the
prospected domain the relative errors do not exceed $10-15\%$. The
BCM procedure provides a reasonable accuracy even for extremely
irregular field of acoustic rays.

\section{Conclusion}

In the present work we have investigated practical capabilities of
the Boundary Control Method applied to acoustic equation on
semiplane. This {\it ab initio} and {\it direct} reconstruction
method does not require any {\it a priory} information about the
recovered parameters of a dynamic system.
To solve inverse problem, the BCM exploits subtle and rigorous properties of the dynamic system.

The BCM reconstruction procedure is simple and efficient,
essentially it involves calculation of quadratures for matrix
elements (double sums) and solving linear systems by standard
LAPACK routines. The method may require additional regularization
(e.g. by Tikhonov) of the linear systems and smoothing (e.g. by
convolution with some kernel) of the Gibbs oscillations arising in
expansions of projections of harmonic functions over wave
solutions. The preparation of synthetic inverse data for BCM is
rather CPU time consuming since we have to solve the direct problem
for $2^8 - 2^{12}$ boundary controls with a good accuracy.

The condition number of matrices $A^\xi$ composed from scalar
products of wave solutions produced by set of boundary controls
quickly grows with the probing time $\xi\in[0, T]$.
This effect is manifestation of ill-posedness of the inverse problem and it
imposes a practical limit on the maximum depth of reconstruction
of speed of sound. To increase the depth of reconstruction we have
to reduce errors in system's reaction data and/or decrease the
number of active boundary controls (the matrix dimension). The use
of {\it sliding} support $\sigma$ allows to decrease the number of
boundary controls without degradation of spatial resolution of the
basis.

The application of BCM to realistic model of speed of sound by
Marmousi has confirmed that the method is able to work in cases
with extremely complicated and irregular field of acoustic rays.

The number and shape of boundary controls determine the spatial
resolution of the reconstruction procedure. The BCM demonstrates
ability to work with low number of boundary controls -- in such
case it recovers an `averaged' profile that can be further used
as a `zero order' starting approximation for high resolution
iterative reconstruction methods.

\section{Acknowledgments}

I.~B.~I. was greatly supported by his wife and partly by Volks-Wagen Foundation.
M.~I.~B. was supported by the grant RFBR 14-01-00535A and by Volks-Wagen Foundation.
All authors acknowledge partial support within the contract 2-53-04-SKIF/NIR/16-SPbSU.


\bibliography{article_final}


\section[Figures]{Figures}

\label{lastpage}

\end{document}